\documentclass[11pt]{article}
\pdfoutput=1 
\usepackage{jheppub}
\usepackage[utf8]{inputenc}

\usepackage{graphicx,ulem}
\usepackage{caption,epigraph}
\usepackage{subcaption}
\usepackage{hyperref}
\usepackage[T1]{fontenc}
\usepackage[export]{adjustbox}
\usepackage[table]{xcolor}
\usepackage{amssymb}
\usepackage{pifont}

\newcommand{\be}{\begin{equation}}
\newcommand{\ee}{\end{equation}}
\newcommand{\bea}{\begin{eqnarray}}
\newcommand{\eea}{\end{eqnarray}}

\newcounter{daggerfootnote}
\newcommand*{\daggerfootnote}[1]{%
    \setcounter{daggerfootnote}{\value{footnote}}%
    \renewcommand*{\thefootnote}{\fnsymbol{footnote}}%
    \footnote[2]{#1}%
    \setcounter{footnote}{\value{daggerfootnote}}%
    \renewcommand*{\thefootnote}{\arabic{footnote}}%
    }

\begin{document} 
\begin{flushright}
CPHT-RR017.042023
\end{flushright}

\allowdisplaybreaks
\flushbottom

\title{U(1) quasi-hydrodynamics: Schwinger-Keldysh effective field theory and holography}

\author[\hbar,\nabla]{Matteo Baggioli,}
\author[\infty]{Yanyan Bu,\daggerfootnote{Corresponding Author}}
\author[\otimes]{and Vaios Ziogas}
\affiliation[\infty]{School of Physics, Harbin Institute of Technology, Harbin 150001, China}
\affiliation[\hbar]{Wilczek Quantum Center, School of Physics and Astronomy, Shanghai Jiao Tong University, Shanghai 200240, China}
\affiliation[\nabla]{Shanghai Research Center for Quantum Sciences, Shanghai 201315, China}
\affiliation[\otimes]{CPHT, CNRS, École polytechnique, Institut Polytechnique de Paris, 91120 Palaiseau, France}

\emailAdd{b.matteo@sjtu.edu.cn}
\emailAdd{yybu@hit.edu.cn}
\emailAdd{vaios.ziogas@polytechnique.edu}

\abstract{We study the quasi-hydrodynamics of a system with a softly broken $U(1)$ global symmetry using effective field theory (EFT) and holographic methods. In the gravity side, we consider a holographic Proca model in the limit of small bulk mass, which is responsible for a controllable explicit breaking of the $U(1)$ global symmetry in the boundary field theory. We perform a holographic Schwinger-Keldysh analysis, which allows us to derive the form of the boundary effective action in presence of dissipation. We compare our results with the previously proposed EFT and hydrodynamic theories, and we confirm their validity by computing the low-energy quasi-normal modes spectrum analytically and numerically. Additionally, we derive the broken holographic Ward identity for the $U(1)$ current, and discuss the recently proposed novel transport coefficients for systems with explicitly broken symmetries. The setup considered is expected to serve as a toy model for more realistic situations where quasi-hydrodynamics is at work, such as axial charge relaxation in QCD, spin relaxation in relativistic systems, electric field relaxation in magneto-hydrodynamics, or momentum relaxation in condensed matter systems.\\

}

\maketitle
\flushbottom

\allowdisplaybreaks

\section{Introduction}
\epigraph{Symmetry, as wide or as narrow as you may define its meaning, is one idea by which man through the ages has tried to comprehend and create order, beauty and perfection.}{Hermann Weyl}
The notion of symmetry plays a central role in the theoretical description of physical phenomena. This pivotal role is even more fundamental in the context of effective descriptions, in which the information about microscopic details is not accessible. This is especially the case for effective field theory (EFT) \cite{Penco:2020kvy} and hydrodynamics \cite{Kovtun:2012rj}. Those are low-energy effective descriptions with a finite regime of validity, which are constructed in a perturbative fashion in terms of only the relevant degrees of freedom. In the context of hydrodynamics, the regime of validity is that of late time and large distances (or equivalently small gradients), and the perturbative structure is known as the \textit{gradient expansion}. In general, symmetries not only play a role of guidance in the construction of the theory, but they also strongly constrain the low-energy dynamics and the nature of the physical excitations. The case of non-linearly realized symmetries and Goldstone bosons is emblematic in this sense \cite{Beekman:2019pmi}. As a consequence of Noether's theorem, symmetries imply the presence of conserved quantities, whose dynamics is governed by the so-called \textit{hydrodynamic modes}, protected excitations whose frequency goes to zero in the limit of zero wave-vector. The possibility of separating the hydrodynamic modes from the non-hydrodynamic, and in this sense irrelevant, ones is a fundamental assumption in the construction of hydrodynamics, also known as \textit{separation of scales}.

EFT and hydrodynamics have been widely used in many different physical situations (\textit{e.g.}, \cite{PhysRevA.6.2401,RevModPhys.95.011001,Penco:2020kvy,PhysRevResearch.2.033124,PhysRev.188.898}), and their success is ultimately due to symmetries. Nevertheless, despite Nature preferring symmetry and simplicity, symmetries are very often approximate or softly broken. Several examples can be mentioned. Finite quark masses in QCD explicitly break chiral symmetry giving a small but finite mass to the pions \cite{Burgess:1998ku}. Impurities or disorder explicitly break translational symmetries, pinning the corresponding Goldstone modes (\textit{e.g.}, charge density waves \cite{RevModPhys.60.1129,RevModPhys.95.011001}). The effects of explicitly breaking a certain symmetry are reflected in the appearance of almost conserved quantities, and of relaxing excitations which decay exponentially in time. These modes are not anymore hydrodynamic, in the sense that their frequency is finite even in the limit of zero wave-vector, $k=0$. In particular, the imaginary part of their frequency at $k=0$ determines their relaxation rate, $-\mathrm{Im}[\omega(k=0)]=\Gamma=\tau^{-1}$ (where $\tau$ is the associated relaxation time), while their real part at $k=0$ gives their mass gap. In the time domain, these modes decay exponentially, as $\sim \exp\left(-\Gamma t\right)$. Therefore, their dynamics is completely negligible for times larger than $1/\Gamma$, \textit{i.e.}, in the hydrodynamic limit. On general grounds, we do expect the relaxation rate $\Gamma$ to parameterize the breakdown of the conservation law and therefore the strength of explicit symmetry breaking. Larger explicit breaking implies a larger relaxation rate for the corresponding non-conserved quantity.

\begin{figure}
    \centering
    \includegraphics[width=0.45 \linewidth]{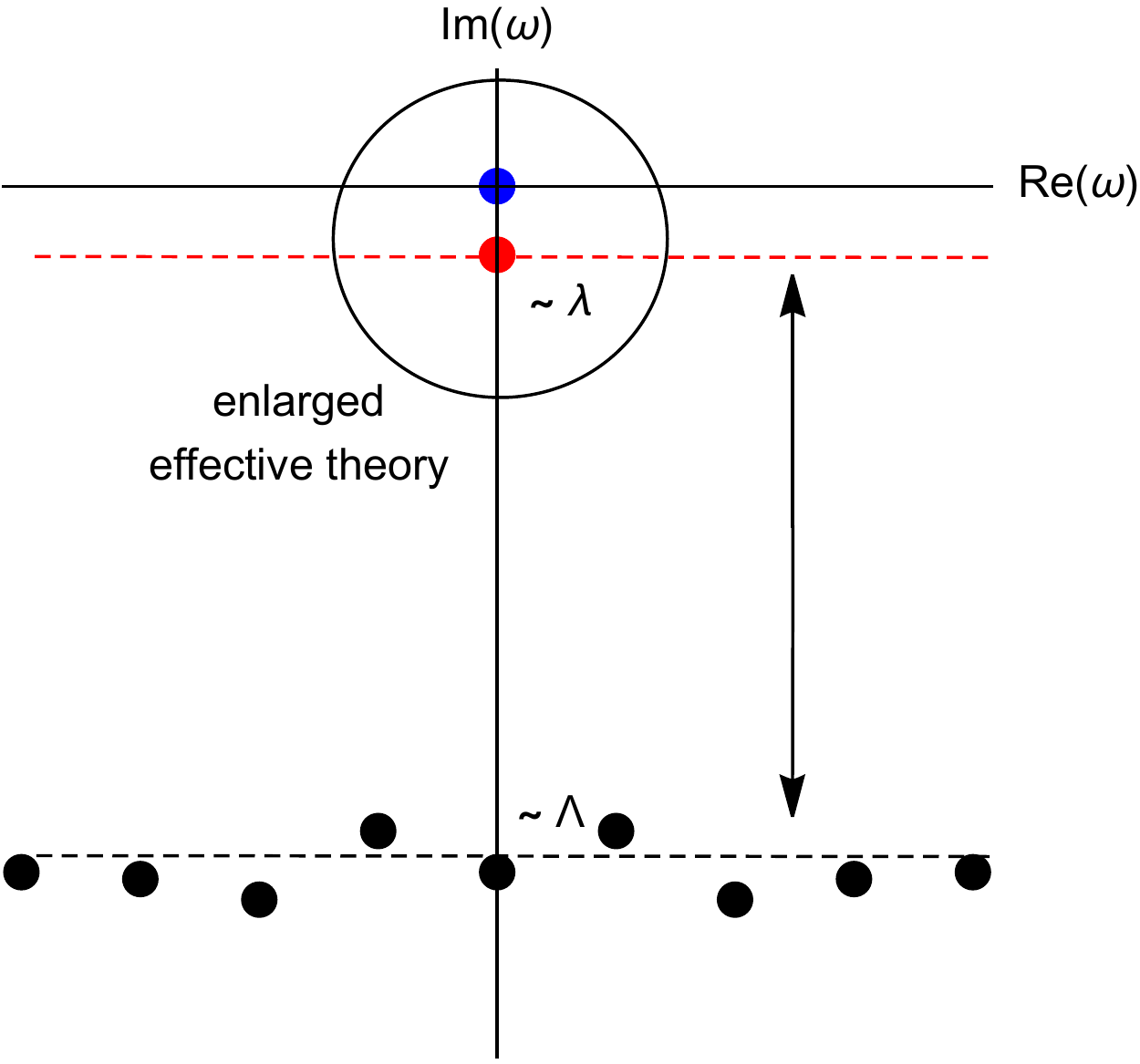}\qquad
    \includegraphics[width=0.45 \linewidth]{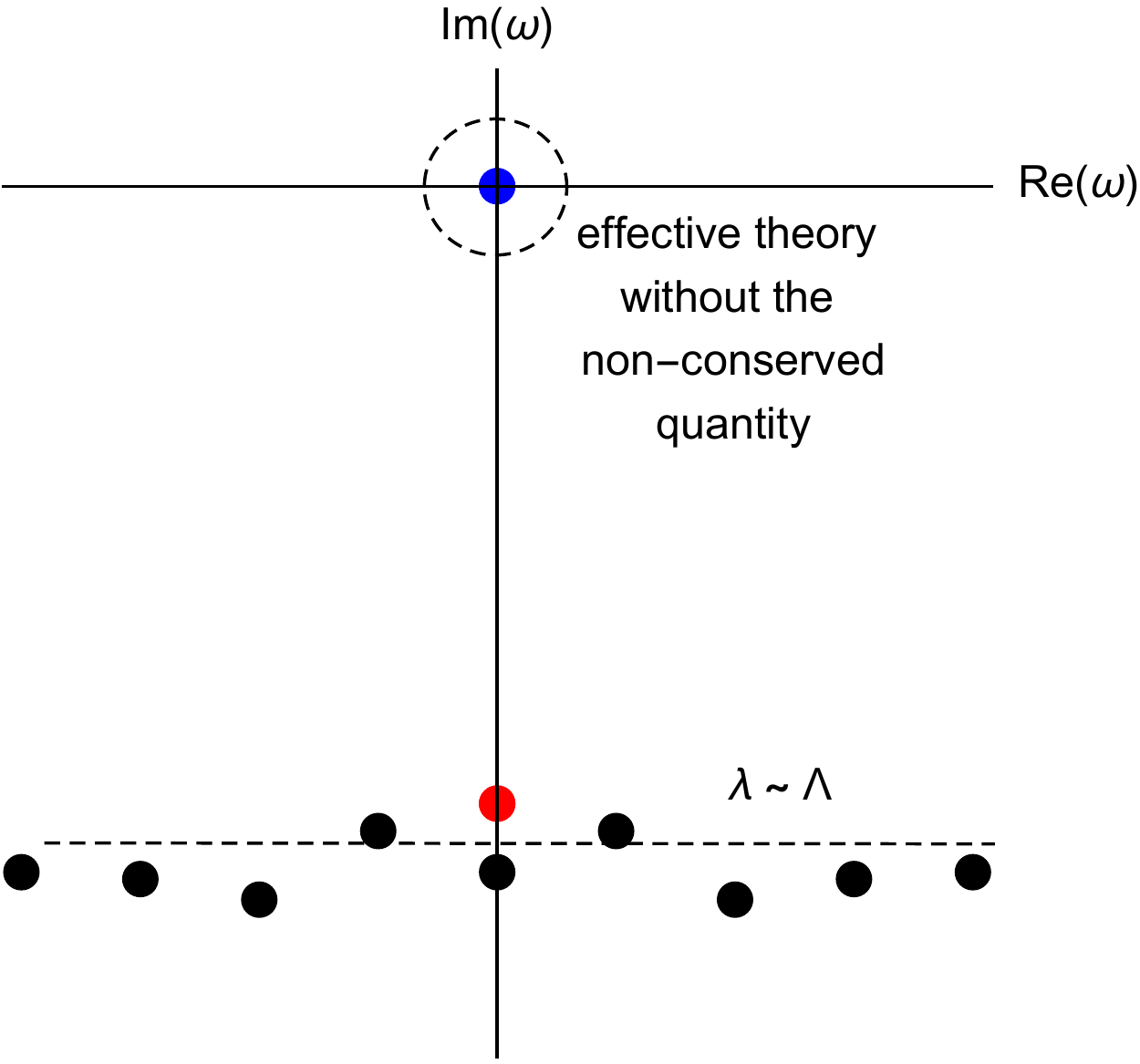}
    \caption{\textbf{Left: }The small breaking regime in which the explicit breaking scale $\lambda$ is much smaller than the microscopic scale $\Lambda$. The red symbol corresponds to the non-conserved quantity, the black ones parameterize the microscopic physics, and the blue ones are the protected hydrodynamic modes. \textbf{Right: }The large breaking regime in which the explicit breaking scale $\lambda$ is of the same order of the microscopic scale $\Lambda$. In this case, no quasi-hydrodynamic description can be pursued. For simplicity, the cartoon is taken at zero wave-vector, $k=0$ and only a single non-hydrodynamic mode, corresponding to a single non-conserved quantity, is considered.}
    \label{fig1}
\end{figure}

Because of these motivations, and also as a fundamental theoretical question, enlarging the EFT and hydrodynamic descriptions in the case of explicitly broken symmetries is of extreme importance, but it encounters foundational problems. In general, this may appear as a challenging or even impossible task. The main difficulty lies in the capability of separating the non-hydrodynamic modes arising due to the broken symmetries from those who are just parameterizing the dynamics of microscopic quantities, which are irrelevant at late time or large scale. A simplification might arise when the explicit breaking strength $\lambda$ can be tuned to be parametrically small, i.e. much smaller than the microscopic scale $\Lambda$, determining the location of the other non-hydrodynamic modes (see Figure \ref{fig1} for a cartoon). In first approximation, we can therefore define the explicit breaking scale to be small whenever $\lambda/\Lambda \ll 1$. In finite temperature systems, the microscopic scale is usually of the order of the temperature $T$, and therefore the small explicit breaking regime is then defined by the condition $\lambda/T \ll 1$. Following the arguments above, one does expect the relaxation rate $\Gamma$ to be also small (compared to $T$) in such a limit. In that case (left panel in Figure \ref{fig1}), one could safely separate the modes associated with the broken symmetries from those related to the microscopic physics. Therefore, one could enlarge the validity of the effective description by incorporating the modes corresponding to the broken symmetry, while still neglecting the rest of the UV ones. Notice that this is possible only because the scale of the non-hydrodynamic modes related to the explicit broken symmetries can be directly controlled by the explicit breaking parameter, and can be made in principle parametrically small. In the context of relativistic hydrodynamics, this extended framework has recently been labelled as \textit{quasi-hydrodynamics} \cite{Grozdanov:2018fic}. On the contrary, in the opposite limit where the explicit breaking parameter is large, $\lambda/\Lambda \gg 1$, such an extended effective description does not exist. Indeed, in such a situation, the non-conserved quantities decay very fast in time, and the imaginary part of the frequency of the corresponding modes is very large, of the same order of the other microscopic excitations (right panel in Figure \ref{fig1}). One has simply to discard the non-conserved quantities from the low-energy effective description and live without them. 

So far, we have mostly discussed the homogeneous dynamics in which the wave-vector of the excitations is taken to be zero, $k=0$. In the complete scenario at finite wave-vector, the discussion about the separation of scales and the various regimes becomes more complicated, since it involves the size the wave-vector $k$ itself. We will discuss this in more detail in Section \ref{sec1}. Notice that a similar quasi-hydrodynamic scenario always appears in the vicinity of a critical point, where critical modes approach the hydrodynamic limit. In that case, the role of the explicit breaking scale is played by the distance from the critical point (\textit{e.g.}, $(T-T_c)$ for a finite temperature critical point). In that context, the extended hydrodynamic description is known as Hydro+ \cite{Stephanov:2017ghc}, and the slowly relaxing non-hydrodynamic mode usually corresponds to the fluctuations of the order parameter \cite{RevModPhys.49.435}. Let us emphasize here that all these extended frameworks are profoundly different from the \textit{generalized hydrodynamics} or mode-coupling theory used in liquid dynamics \cite{boon1991molecular}, in which the non-hydrodynamic modes are not parameterized by a tunable perturbative parameter, but they are just a manifestation of some unknown microscopic physics. Needless to say, the extension of hydrodynamics beyond its original regime of validity has attracted a lot of interest in the context of classical fluids, with the appearance of interesting phenomena such as the cutoff wave-vector for propagating shear waves \cite{BAGGIOLI20201}, and the positive sound dispersion for longitudinal sound \cite{ruocco2008history}. Recently, the important role of non-hydrodynamic modes has been also appreciated in the context of relativistic hydrodynamics and heavy-ion collisions \cite{Brewer:2022ifw,Ke:2022tqf}.

Apart from the conceptual difficulties, writing down an enlarged effective description in the presence of broken symmetries comes with additional operational difficulties. In particular, allowing for all the possible effects of this soft explicit breaking might be cumbersome. On top of that, the structure of the gradient expansion and the order therein of the various quantities (frequency, wave-vector, explicit breaking scale) become very subtle. At the same time, the role of frame redefinitions is not clearly established. In the past, these obstacles have created a lot of confusion and contrasting results (some of which still exist). In this direction, holography has emerged as an extremely useful and effective tool to help with the construction of hydrodynamics and effective field theories with broken symmetries (see for example the case of charge density waves, \cite{PhysRevA.6.2401,PhysRevB.96.195128,Ammon:2019apj,Donos:2019hpp,Armas:2019sbe,Ammon:2020xyv}). The merits of holography are twofold. On the one hand, holography provides the full structure of the hydrodynamics or EFT expansion at all orders in the perturbative parameters (frequency and wave-vector, for example). There is no space left for missing terms or neglected effects. On the other hand, holography gives a fully microscopic description able to concretely predict the numerical values of all the coefficients involved in the hydrodynamic or EFT perturbative expansion. In the case of systems with broken symmetries, holography has another big advantage. The holographic description is valid not only in the limit of small explicit breaking but at every order in its effects. This implies that holography can also play an important role to ascertain and quantitatively predict the regime of validity of the extended effective description described above, \textit{i.e.}, how far quasi-hydrodynamics can go.

One of the main critiques towards the so-called bottom-up holography pertains the complete (or almost) ignorance about the structure of the dual field theory description. This could be a well motivated criticism, even though in several contexts (\textit{e.g.}, holographic superfluids \cite{Arean:2021tks}, holographic systems with broken translations \cite{Ammon:2020xyv}, magneto-hydrodynamics \cite{Ahn:2022azl}, etc.), a quantitative match between the EFT/hydrodynamic description of the dual field theory and the holographic results has been shown. On top of several numerical checks, in recent years, the situation has drastically changed thanks to the so-called holographic Schwinger-Keldysh (SK) techniques \cite{Glorioso:2018mmw,Crossley:2015tka,Haehl:2015foa}. This new tool allows for a direct and explicit determination of the dual effective theory including all its dissipative terms, and is particularly useful as it provides an (symmetry-based) \textit{action principle} for hydrodynamics,\footnote{See \cite{Grozdanov:2013dba,Kovtun:2014hpa} for some earlier attempts.} and directly computes real-time observables. Using the aforementioned methods, one is able to obtain the full SK effective field theory description \cite{kamenev2011field,Liu:2018kfw} of a given holographic model directly from its gravitational bulk dynamics. These techniques have been rapidly growing and have already been applied to many holographic systems \cite{Chakrabarty:2019aeu,Jana:2020vyx,Loganayagam:2020iol,Ghosh:2020lel,Bu:2020jfo,Bu:2021clf,He:2021jna,Bu:2021jlp,Bu:2022esd,He:2022deg,Bu:2021jlp,Ghosh:2022fyo,He:2022jnc,Bu:2022oty,Loganayagam:2022zmq,Pantelidou:2022ftm}.

Hydrodynamics and effective field theory with explicitly broken symmetries have been recently considered in the context of translational symmetry \cite{Delacretaz:2017zxd,Armas:2021vku,Baggioli:2022pyb,Baggioli:2020haa,Delacretaz:2021qqu,Landry:2020ire,Baggioli:2020nay}, $U(1)$ symmetry \cite{Landry:2020tbh,Ammon:2021pyz,Delacretaz:2021qqu}, non-Abelian symmetries \cite{Grossi:2020ezz,Grossi:2021gqi,Cao:2022csq} and also higher-form symmetries \cite{Landry:2021kko,Baggioli:2021ntj}. Holography has played a fundamental role both in verifying but also falsifying the proposed effective descriptions (see for example \cite{Ammon:2019apj}). In this work, we will consider the simplest scenario possible for studying the effective description of systems with explicitly broken symmetries and its quasi-hydrodynamic regime. More specifically, using both SK field theory and holographic SK techniques, we will study in detail a system which explicitly breaks a global $U(1)$ symmetry in the limit of small explicit breaking, which can serve as a toy model for the more complicated cases of translations and non-Abelian symmetries. For simplicity, we will also work in the probe limit in which the dynamics of the stress tensor (\textit{i.e.}, temperature and momentum fluctuations) is kept frozen.

\subsection{U(1) quasi-hydrodynamics}\label{sec:u1intro}
Let us consider a finite temperature system which possesses a global $U(1)$ symmetry. The latter is associated to the conservation of a charge $\mathcal{Q}$ (\textit{e.g.}, electric charge) and of a current density $J^\mu$:
\begin{equation}\label{eq1}
    \partial_\mu J^\mu=0,
\end{equation}
which, in absence of sources, arises as the Ward identity for the aforementioned $U(1)$ symmetry. In the probe limit, neglecting the dynamics of the stress tensor, the only relevant hydrodynamic degree of freedom at low energy corresponds to the fluctuations of the charge density $\delta \rho(t,
    \Vec{x})$. At lowest order in the gradient expansion, or equivalently in the wave-vector $k$, charge fluctuations obey a diffusive equation given by:
\begin{equation}
    \left(\partial_t + D \nabla^2\right)\delta \rho(t,
    \Vec{x})=0,
\end{equation}
which is a direct consequence of the macroscopic Fick's law. In the context of hydrodynamics, the latter manifests itself in the so-called constitutive relation for the current $J^\mu$. Finally, still within the effective description, one can derive that the diffusion constant $D$ is given by:
\begin{equation}
    D=\frac{\sigma}{\chi_{\rho\rho}}\,,\quad \text{with}\quad \chi_{\rho\rho}\equiv \frac{\partial \rho}{\partial \mu},
\end{equation}
with $\sigma$ being the electric conductivity, $\chi_{\rho\rho}$ the charge susceptibility and $\mu$ the chemical potential. What we have just described is what one observes by dissolving color in a glass of water. For a formal derivation using standard hydrodynamics see \cite{Kovtun:2012rj}; for a derivation using SK techniques see \cite{Liu:2018kfw}.

What happens when we do break the $U(1)$ global symmetry explicitly? The total charge is not anymore conserved and charge fluctuations decay exponentially in time at a rate $\Gamma$, proportional to the amount of explicit breaking (at least, for small explicit breaking). If this rate is too large, charge is gone before we have the time to actually see it, meaning that the late time description of such a system does not need to take into account the dynamics of charge fluctuations at all. What happens on the contrary if the breaking is very soft and the relaxation rate $\Gamma$ small? Then, one expects that the conservation equation \eqref{eq1} can be somehow retained, but modified by the addition of new terms in the r.h.s. which parameterize the amount of non-conservation:
\begin{equation}\label{eqgen}
    \partial_\mu J^\mu= \includegraphics[height=0.5cm, valign=c]{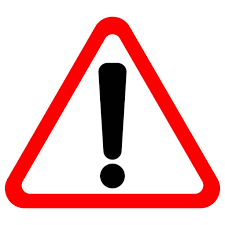}\,\,.
\end{equation}

Recall that the general philosophy of hydrodynamics is that we can expand any operator in terms of gradients of the slow degrees of freedom, and this will hold in particular for the operators appearing in the r.h.s. of \eqref{eqgen}. When the only relevant degree of freedom is the charge density $\rho$, the hydrodynamic expansion will give rise to (in a generally covariant form)\footnote{See \cite{Abbasi:2018qzw,Amoretti:2023vhe} for further discussions on relaxed hydrodynamics and covariance.}
\begin{equation}\label{eq2}
    \partial_\mu J^\mu= \Gamma \,u_\mu J^\mu +\cdots\,,
\end{equation}
at leading order, where $u_\mu$ is the four-velocity vector, and $\Gamma$ is a dissipative transport coefficient from this point of view. In other words, flowing to the IR of the deformed QFT is expected to lead to the effective Ward identity Eq. \eqref{eq2}. Alternatively, one can postulate Eq. \eqref{eq2} directly as the effective description of the system. This approach is quite common from the EFT point of view, and is inherently agnostic of the microscopic details of the underlying QFT; for instance, \eqref{eq2} simply encodes that the current operator $J^\mu$ acquires an anomalous dimension in the IR.

Recently, a lot of activity has focused on the pinned or relaxed phase of superfluids, in which the $U(1)$ is also spontaneously broken, on top of the soft explicit breaking. The hydrodynamics of such systems has been derived in the probe limit in \cite{Ammon:2021pyz,Delacretaz:2021qqu,Armas:2021vku}. Holographic models have been employed in \cite{Donos:2021pkk,Ammon:2021pyz} in order to analyze this problem in more detail. Ref.~\cite{Donos:2021pkk} took a standard approach by breaking the global $U(1)$ symmetry of the dual field theory with explicit sources for charged scalar operators. On the contrary, Ref.~\cite{Ammon:2021pyz} considered, in addition to the scenario just mentioned, a different one in which the global $U(1)$ symmetry of the dual field theory is explicitly broken by the presence of a mass term for the bulk gauge field. This kind of gravitational action is known as Proca theory, and its use has been inspired by the applications of the holographic correspondence to mimic the dynamics of non-conserved axial charge \cite{Klebanov:2002gr,Gursoy:2014ela,Jimenez-Alba:2014iia,Jimenez-Alba:2015awa,Iatrakis:2015fma,Iatrakis:2014dka,Bigazzi:2018ulg,Rai:2023nxe}.

Despite the compatibility of the numerical results of \cite{Ammon:2021pyz} with the hydrodynamic description and with the broken Ward identity in Eq.~\eqref{eq2}, several points of the analysis therein have not been clarified. First, there is a lot of confusion in the literature about the broken Ward identity in Eq.~\eqref{eq2} for the holographic Proca model. Additionally, in presence of a finite charge density state, it is unclear how one can even define thermal equilibrium in presence of a softly broken $U(1)$ symmetry and a non-conservation equation as in Eq.~\eqref{eq2}. At most, one expects the possibility of creating a steady state by balancing the charge relaxation term in the r.h.s. of Eq.~\eqref{eq2} with some external source and/or driving (see for example \cite{Amoretti:2022ovc}). Finally, Ref.~\cite{Armas:2021vku} identified additional transport coefficients which appear when the $U(1)$ is pseudo-spontaneously broken, and it would be interesting to gain a better understanding of their origin.\\

In this manuscript, we consider the holographic Proca model and we study in detail its low-energy dynamics using holographic SK techniques. We see the emergence of the SK effective action and we explicitly compute many of its coefficients in terms of bulk quantities. As a corollary, we derive the Ward identity in Eq.\eqref{eq2}. We also find analogous transport coefficients to the ones introduced in \cite{Armas:2021vku}, albeit they appear at higher order in the explicit breaking scale. Finally, we derive the dispersion relation of the pseudo-diffusive hydrodynamic mode, and observe that, to leading order, its diffusion constant does not receive corrections. Apart from these investigations, we aim to pave the way towards extending the holographic Scwhinger-Keldysh techniques in presence of explicitly broken symmetries into the quasi-hydrodynamic regime.\\

The manuscript is organized as follows. In Section \ref{sec1}, we review and extend the EFT description for systems with broken $U(1)$ symmetry; in Section \ref{sec2} we perform the Schwinger-Keldysh analysis for the holographic Proca model; in Section \ref{sec3} we present the main results of our work and the lessons learned, and finally, in Section \ref{sec4}, we conclude with an outlook and some perspectives for the future. The details of the holographic Schwinger-Keldysh computations appear in Appendix \ref{app1}.

\section{Schwinger-Keldysh effective field theory for relaxed U(1) diffusion} \label{sec1}

The EFT for a single conserved $U(1)$ charge and its diffusive dynamics  has been constructed in \cite{Crossley:2015evo}, and later confirmed by holographic methods in \cite{Glorioso:2018mmw,deBoer:2018qqm,Bu:2020jfo}. Here, we will extend the construction of \cite{Crossley:2015evo} to the situation in which a soft explicit breaking of the global $U(1)$ symmetry is introduced. Our building blocks will be the gauge-invariant quantities $B_{s\mu}$ and $\vartheta_s$
\begin{align}
    B_{s\mu} \equiv A_{s\mu} + \partial_\mu \phi_s, \qquad \qquad \vartheta_s = \theta_{bs} + \phi_s,
\end{align}
where $s=1$ ($s=2$) corresponds to the upper (lower) branch of the SK closed time path, $\phi_s$ is the dynamical field, $A_{s\mu}$ is an external gauge field coupling to the $U(1)$ current, and $\theta_{bs}$ is an external scalar source introduced in order to artificially restore the $U(1)$ symmetry \cite{Armas:2021vku}. The terms constructed from the building block $\vartheta_s$ will be responsible for the non-conservation of the global $U(1)$ current, \textit{cfr.} Eq.\eqref{eq2}. In general, it is more convenient to work with the Keldysh retarded-advanced basis
\begin{align}
    B_{r\mu} = \frac{1}{2} (B_{1\mu} + B_{2\mu}), \qquad \vartheta_r = \frac{1}{2} (\vartheta_1 + \vartheta_2), \qquad B_{a\mu} = B_{1\mu} - B_{2\mu}, \qquad \vartheta_a = \vartheta_1 - \vartheta_2\,.
\end{align}
The EFT action can be organized in a perturbative hydrodynamic expansion in derivatives as 
\begin{align}\label{eq:seffL1L2}
    S_{\text{eff}}= \int d^4x \left( \mathcal L_1 + \mathcal L_2 \right),
\end{align}
where:
\begin{align}
    \mathcal{L}_1&=g_{0} B_{a0}B_{r0}+g_{1} B_{a0}\partial_0 B_{r0} + f_{0}B_{ri}B_{ai} + f_{1}B_{ai}\partial_0 B_{ri} + h_{0}B_{a0}\partial_i B_{ri}  \nonumber \\
    &+ h_{1}B_{ai}\partial_i B_{r0} +\Gamma_0 \vartheta_a\vartheta_r +\tilde \Gamma \vartheta_a\partial_0\vartheta_r +\Gamma_1 \vartheta_a\partial_0^2\vartheta_r +\Gamma_2 \vartheta_a\partial_i^2\vartheta_r +c_0 \vartheta_a B_{r0} \nonumber \\
    & +c_1 \vartheta_a \partial_0 B_{r0} +c_2 \vartheta_a \partial_i B_{ri} +d_0 \vartheta_r B_{a0} +d_1 \partial_0 \vartheta_r B_{a0} +d_2 \partial_i \vartheta_r B_{ai} \nonumber \\
    & + \frac{1}{2}w_{0} B_{a0}^2+w_{1} B_{a0}\partial_i B_{ai} + \frac{1}{2}w_{2}B_{ai}^2 + \frac{1}{2}w_{3}\vartheta_a^2 +\frac{1}{2}w_4 \vartheta_a\partial_i^2\vartheta_a + \frac{1}{2} w_5\vartheta_a \partial_0^2 \vartheta_a \nonumber \\
    &+w_6 \vartheta_a B_{a0} +w_7 \vartheta_a \partial_0 B_{a0} +w_8 \vartheta_a \partial_i B_{ai},  \label{L1}\\
    \mathcal L_2 & =  c_3 \vartheta_a \partial_0^2 B_{r0} + c_4 \vartheta_a \partial_0 \partial_i B_{ri} + c_5 \vartheta_a \partial_i^2 B_{r0}  + d_3 \partial_0^2 \vartheta_r B_{a0} + d_4 \partial_0 \partial_i \vartheta_r B_{ai}  \nonumber \\
    &+d_5 \partial_i^2 \vartheta_r B_{a0}  + \Gamma_3 \vartheta_a \partial_0 \partial_i^2 \vartheta_r + \Gamma_4 \vartheta_a \partial_0^3 \vartheta_r + g_2 B_{a0} \partial_0^2 B_{r0} + g_3 B_{a0} \partial_i^2 B_{r0}  \nonumber \\
    & + f_2 B_{ai} \partial_i \partial_k B_{rk} + f_3 B_{ai} \partial_j^2 B_{ri} + h_2 B_{a0} \partial_0 \partial_i B_{ri} + h_3 B_{ai} \partial_0 \partial_i B_{r0} \nonumber \\
    & + \Gamma_5 \vartheta_a \partial_0^4 \vartheta_r + \Gamma_6 \vartheta_a \partial_0^2 \partial_i^2 \vartheta_r + \Gamma_7 \vartheta_a \partial_i^2 \partial_j^2 \vartheta_r + c_6 \vartheta_a \partial_0^3 B_{r0} + c_7 \vartheta_a \partial_0 \partial_i^2 B_{r0} \nonumber \\
    & + c_8 \vartheta_a \partial_0^2 \partial_i B_{ri} + c_9 \vartheta_a \partial_j^2 \partial_i B_{ri} + d_6 B_{a0} \partial_0^3 \vartheta_r + d_7 B_{a0} \partial_0 \partial_i^2 \vartheta_r + d_8 B_{ai} \partial_0^2 \partial_i \vartheta_r \nonumber \\
    & + d_9 B_{ai} \partial_k^2 \partial_i \vartheta_r + w_9 \vartheta_a \partial_0^2 B_{a0} + w_{10} \vartheta_a \partial_i^2 B_{a0} + w_{11} \vartheta_a \partial_0 \partial_i B_{ai} + \frac{1}{2} w_{12} \vartheta_a \partial_0 \partial_i^2 \vartheta_a\nonumber \\
    & + \frac{1}{2} w_{13} \vartheta_a \partial_0^3 \vartheta_a,  \label{L2}
\end{align}
where we have imposed rotational invariance along the spatial coordinates. Setting $\vartheta_r = \vartheta_a =0$, \eqref{L1} and \eqref{L2} reduce to the effective Lagrangian for single conserved $U(1)$ charge, as in \cite{Crossley:2015evo}. Thus, the coefficients in \eqref{L1} and \eqref{L2} generally scale as
\begin{align}\label{eq:derivative_scheme}
    g_{0-3},\, f_{0-3},\, h_{0-3}, \, w_{0-2} \sim \mathcal{O}(m^0), \qquad \tilde \Gamma, \, \Gamma_{0-7}, \,  c_{0-9}, d_{0-9}, \, w_{3-13} \sim \mathcal{O}(m^2),
\end{align}
as a function of the parameter $m$, which is the scale characterizing the soft explicit breaking of the $U(1)$ symmetry. Here, we have clearly assumed that $m/T \ll 1$, and that therefore such an expansion is legitimate. This scaling makes the terms $\mathcal{L}_1$ leading compared to those in $\mathcal{L}_2$, in an $\mathcal{O}(\partial,m^2)$ expansion, justifying the splitting made in \eqref{eq:seffL1L2}.

Importantly, in this work, we will only consider a (soft) explicit breaking of the global $U(1)$ symmetry and not an additional spontaneous breaking of the latter, as in \cite{Ammon:2021pyz,Delacretaz:2021qqu,Armas:2021vku}. This will have two crucial consequences. First, in absence of spontaneous breaking, chemical shifts are an exact symmetry of the system which the SK functional should obey. Second, the leading scaling of several coefficients with respect to the explicit breaking scale will be profoundly different in the purely explicit case compared to the pseudo-spontaneous one. We will return on this point later in the manuscript.

We turn to the various constraints among the coefficients appearing in \eqref{L1} and \eqref{L2}, which are imposed by the general rules of the Scwhinger-Keldysh effective field theory (SK EFT) \cite{Crossley:2015evo,Glorioso:2017fpd}.

Here, we provide a list of them and the corresponding constraints.\\

\noindent $\bullet$ \emph{$Z_2$-reflection symmetry}:
\begin{align}
   \left( S_{eff}[B_{r\mu}, \vartheta_r; B_{a\mu}, \vartheta_a] \right)^* = - S_{eff}[B_{r\mu}, \vartheta_r; - B_{a\mu}, - \vartheta_a], \label{Z2_symmetry}
\end{align}
which directly follows from the SK formalism. This symmetry implies that the coefficient for any term with an even number of $a$-variables shall be purely imaginary, while the coefficient for any term with an odd number of $a$-variables must be real. Thus, all the $w_i$'s are purely imaginary, and all the other coefficients are real.\\

\noindent $\bullet$ \emph{Chemical shift symmetry}. This symmetry simply reflects the redundancy of the description in labelling the $U(1)$ phase of each local fluid element. It defines the normal phase of a fluid, and is broken only when the $U(1)$ symmetry is spontaneously broken \cite{Crossley:2015evo}. Then, the EFT action is invariant under a diagonal time-independent shift
\begin{align}
    &\phi_r \to \phi_r + \lambda(\vec x), \qquad \qquad \phi_a \to \phi_a \nonumber \\
    \Rightarrow \,& B_{ri} \to B_{ri} + \partial_i \lambda(\vec x), \qquad  \vartheta_r \to \vartheta_r + \lambda(\vec x), \qquad B_{r0}, \, B_{a\mu}, \, \vartheta_a ~ {\rm unchanged},    \label{chemical_shift}
\end{align}
which leads to
\begin{align}
    &d_2 = -f_0, \qquad h_0 =0, \qquad c_2 = -\Gamma_2, \qquad \Gamma_0 =0, \qquad d_0 =0, \nonumber \\
    &d_5 =- h_0, \qquad f_2+ f_3 = -d_9, \qquad \Gamma_7 = -c_9.
\end{align}

\noindent $\bullet$ \emph{Onsager relations}. The Onsager relations follow from the symmetry properties of the retarded (or advanced) correlation functions under a change of the ordering of operators \cite{Crossley:2015evo}. Imposing them, we find
\begin{align}
   & h_0 = h_1 \qquad \quad  h_2 = h_3, \nonumber \\
   & c_0  = - d_0, \qquad   c_1 = - d_1, \qquad c_2 = -d_2, \qquad c_3 = - d_3, \qquad c_4 = -d_4, \nonumber \\
   &  c_5 = -d_5, \qquad  c_6 =- d_6, \qquad c_7 = - d_7, \qquad c_8 = - d_8, \qquad c_9 = - d_9.  \label{Onsager_relation}
\end{align}

\noindent $\bullet$ \emph{Dynamical KMS symmetry}. When the system is in a thermal equilibrium state, the KMS condition sets important constraints on the generating functional $W = i \log Z$, with $Z = \int [D \phi_r] [D \phi_a] {\rm exp}\left( i S_{eff}\right)$ being the partition function. At quadratic level, as considered in this work, the constraint is simply the celebrated fluctuation-dissipation theorem. Within the SK EFT, the KMS condition is guaranteed by imposing that the EFT action satisfies the dynamical KMS symmetry \cite{Glorioso:2017fpd}
\begin{align}
    S_{eff}[B_{1\mu}, \vartheta_1;  B_{2\mu}, \vartheta_2] = S_{eff}[\tilde B_{1\mu}, \tilde \vartheta_1;  \tilde B_{2\mu}, \tilde \vartheta_2], \label{dynamical_KMS}
\end{align}
where, in the classical statistical limit \cite{Glorioso:2017fpd}, 
\begin{align}
    &\tilde B_{r\mu}(-v, -\vec x) = (-1)^{\eta_\mu} B_{r\mu}(v, \vec x), \qquad  \tilde B_{a\mu}(-v, -\vec x) = (-1)^{\eta_\mu} \left[ B_{a\mu}(v, \vec x) + i\beta \partial_0 B_{r\mu} (v, \vec x) \right], \nonumber \\
    &\tilde \vartheta_{r\mu}(-v, -\vec x) = (-1)^{\eta_\vartheta} \vartheta_{r\mu}(v, \vec x), \qquad  \tilde \vartheta_{a\mu}(-v, -\vec x) = (-1)^{\eta_\mu} \left[ \vartheta_{a\mu}(v, \vec x) + i\beta \partial_0 \vartheta_{r\mu} (v, \vec x) \right]. \label{classical_KMS_transform}
\end{align}
Here, $(-1)^{\eta_\mu}$ and $(-1)^{\eta_\vartheta}$ are the eigenvalues of a discrete symmetry transformation $\Theta$
(containing time-reversal $\mathcal T$) acting on $B_\mu$ and $\vartheta$, respectively. We use $v$ to denote the time coordinate, and $\beta\equiv 1/T$ is the inverse temperature. Physically, the dynamical KMS symmetry
\eqref{dynamical_KMS} amounts to imposing microscopic time-reversibility and local equilibrium \cite{Glorioso:2017fpd}. Indeed, by taking the classical statistical limit \eqref{classical_KMS_transform}, the EFT becomes a classical statistical theory, in which only statistical thermal fluctuations survive. Eventually, after imposing \eqref{dynamical_KMS} and \eqref{classical_KMS_transform}, we find
\begin{align}\label{eq:KMS_conditions}
    &g_1 = \frac{1}{2} i\beta w_0, \qquad  f_1 = \frac{1}{2} i \beta w_2, \qquad h_0 =0, \qquad h_1 =0, \qquad \tilde \Gamma = \frac{1}{2} i\beta w_3, \nonumber \\
    &c_0 =0, \qquad d_0=0,  \qquad w_1 =0, \qquad w_6 =0, \qquad c_3 = -d_3 = \frac{1}{2} i\beta w_7, \nonumber \\
    & c_4 = -d_4 = \frac{1}{2} i\beta w_8, \quad c_5 =0, \qquad d_5 =0, \qquad \Gamma_3 =  \frac{1}{2} i \beta w_4, \qquad \Gamma_4 =  \frac{1}{2} i \beta w_5 ,\nonumber \\
    &w_9 = w_{10} = w_{11} = w_{12} = w_{13} =0.
\end{align}
It is interesting to note that in this quasi-hydrodynamic setup, not all of the Onsager relations follow from the dynamical KMS conditions, in contrast to fluids with preserved or spontaneously broken symmetries \cite{Glorioso:2017fpd}.

\noindent $\bullet$ \emph{Well-defined path integral}. Recall that some coefficients in the EFT action are purely imaginary. Then, for the path integral based on $S_{eff}$ to be well-defined, we shall impose
\begin{align}
    {\rm Im}(w_0) \geq 0, \qquad {\rm Im}(w_2) \geq 0, \qquad {\rm Im}(w_3) \geq 0, \quad ...
\end{align}
where the second requirement, combined with KMS relations, indicates that the diffusion constant and conductivity (\textit{i.e.}, the dissipative coefficients) are non-negative. At quadratic level, this is equivalent to the positivity of entropy production, as commonly imposed in hydrodynamics.

In Section \ref{sec2}, through a holographic study, we will confirm the structure of $\mathcal L_1$ in \eqref{L1}, as well as various constraints set by symmetry rules. Moreover, we will compute all the coefficients in \eqref{L1} within the holographic model. $\mathcal L_2$ in \eqref{L2} contains higher order derivative terms, which are crucial in obtaining a correct prediction for the dispersion relation of the low energy modes up to order $\mathcal O(m^2 k^2)$.

\section{Schwinger-Keldysh effective field theory from holography}\label{sec2}

In this section, we apply the holographic prescription for the SK closed time path \cite{Glorioso:2018mmw} to the holographic Proca model. In addition to a holographic derivation of the EFT action written down in section \ref{sec1}, we also compute the unknown coefficients in the EFT action. For simplicity, we will work in the probe limit and consider zero background charge density.

Let us consider the $5$-dimensional Proca theory using the St\"uckelberg formalism\footnote{Alternatively, we could use a more standard Maxwell action coupled to a charged scalar field with a non zero source at the boundary. For the appropriate choice of the scalar conformal dimension, this would  lead to a similar asymptotic analysis as the one performed in section \ref{holo_renor}, and a similar structure for the SK EFT action. This comparison is for example discussed in Section 4.1 of \cite{Jimenez-Alba:2015awa}.}
    \begin{align}
    S_{\rm bulk}= \int d^5x \sqrt{-g} \left[ - \frac{1}{4}F_{MN} F^{MN} - \frac{m^2}{2} (C_M - \nabla_M \theta)(C^M - \nabla^M \theta) \right], \label{Stueckelberg_action}
    \end{align}
which, in the unitary gauge $\theta=0$, reduces to the ``standard'' Proca massive vector theory. The AdS radius has been set to unity $L_{AdS}=1$, and can be reinstated from dimensional analysis. While the $U(1)$ gauge field $C_M$ is massive, the St\"uckelberg theory \eqref{Stueckelberg_action} still enjoys a gauge symmetry
    \begin{align}
        C_M \to C_M + \nabla_M \xi, \qquad  \theta \to \theta + \xi ,\label{gauge_symmetry}
    \end{align}
which will play a crucial role in the subsequent analysis. The bulk action \eqref{Stueckelberg_action} shall be supplemented with a counter-term action $S_{\rm ct}$ which will be specified in section \ref{holo_renor}.

\begin{figure}[htbp]
\centering
\includegraphics[width=0.8\textwidth]{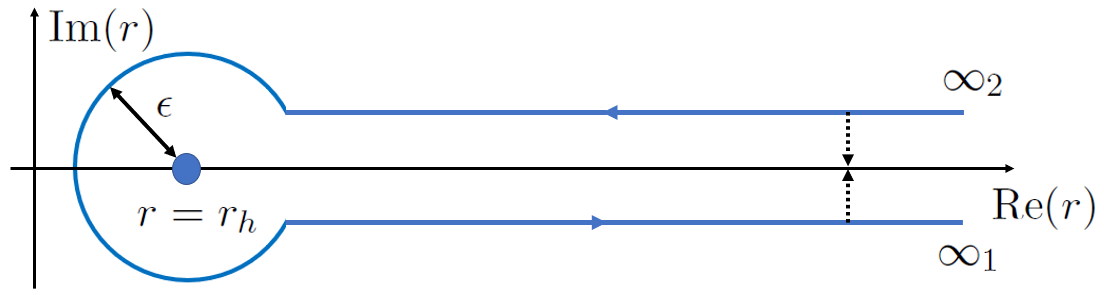}
\caption{The holographic prescription for the Schwinger-Keldysh closed time path \cite{Glorioso:2018mmw}: complexified radial coordinate and analytical continuation around the horizon $r_h$.}
\label{holographic_SK_contour}
\end{figure}

In the ingoing EF coordinate system, the metric for the fixed background geometry is given by the AdS black brane
\begin{align}\label{eq:ads_bb}
    ds^2 = g_{MN} dx^M dx^N = 2drdv - r^2 f(r) dv^2 + r^2\delta_{ij} dx^i dx^j,
\end{align}
with $f(r) = 1-r_h^4/r^4$. In order to construct Schwinger-Keldysh EFT on the boundary, we adopt the holographic prescription of \cite{Glorioso:2018mmw}. Then, the radial coordinate $r$ varies on the contour in Figure \ref{holographic_SK_contour}.

Latin indices, $M,N,\cdots$, are used to denote the $5$-dimensional spacetime coordinates, Greek indices, $\mu, \nu, \cdots$, indicate the $4$-dimensional boundary ones.

\subsection{Holographic dictionary revisited} \label{holo_dictionary}

In order to derive the effective action from a bulk theory in AdS space, we need to consider the holographic Wilsonian renormalization group, as outlined in \cite{Crossley:2015tka} for pure gravity. In \cite{Bu:2020jfo,Bu:2021clf}, this analysis was adopted to the examples of a probe Maxwell theory and scalar QED, respectively, in a fixed AdS black brane. Here, we carry out such an analysis for the St\"uckelberg theory in Eq.~\eqref{Stueckelberg_action}.

The starting point is the Gubser-Polyakov-Klebanov-Witten (GPKW) master rule \cite{Witten:1998qj,Gubser:1998bc}
\begin{align}
    Z_{\rm AdS} = Z_{\rm CFT},
\end{align}
where the partition function $Z_{\rm CFT}$ is expressed as a path integral over the low energy modes collectively denoted by $\phi$,
\begin{align}
    Z_{\rm CFT} = \int [D \phi] e^{i S_{eff}[\phi]}.
\end{align}
Here, $S_{eff}[\phi]$ is the desired effective action to be derived from the holographic model. On the other hand, the AdS partition function is given by
\begin{align}
    Z_{\rm AdS} = \int [DC_M^\prime] [D \theta^\prime] e^{i S_{\rm bulk}[C_M^\prime, \theta^\prime] + i S_{\rm ct}}, \label{Z_AdS1}
\end{align}
where a primed field means no gauge-fixing has yet been imposed.

When $m\neq 0$, the asymptotic behavior for the bulk vector field gets modified dramatically. Accordingly, we set the boundary conditions for $C_M^\prime, \theta^\prime$ as follows (\textit{i.e.}, fixing the non-normalizable modes):
\begin{align}
    r^{-\Delta}C_\mu^\prime(r=\infty,x^\alpha)= A_\mu(x^\alpha), \qquad \theta^\prime (r=\infty,x^\alpha) = \theta_b(x^\alpha) \label{boundary_condition1},
\end{align}
where $\Delta = -1 + \sqrt{1+m^2}$ is the conformal dimension of the boundary $U(1)$ current. 

The bulk gauge symmetry in Eq.~\eqref{gauge_symmetry} allows to bring an arbitrary field configuration $(C_M^\prime, \theta^\prime)$ to a desired one $(C_M, \theta)$ by means of the following transformation
\begin{align}
    C_M^\prime \to C_M = C_M^\prime + \partial_M \xi, \qquad \theta^\prime \to \theta = \theta^\prime + \xi . \label{gauge_transformation1}
\end{align}
Immediately, under \eqref{gauge_symmetry}, the boundary values of the bulk fields transform as
\begin{align}
   r^{-\Delta} C_\mu(r=\infty,x^\alpha)= A_\mu + \partial_\mu \phi \equiv B_\mu, \qquad \theta(r=\infty,x^\alpha)= \theta_b + \phi \equiv \vartheta , \label{AdS_conditions}
\end{align}
where
\begin{equation}
    \phi=r^{-\Delta}\xi(r=\infty,x^\alpha)\,.
\end{equation}
 For an infinitesimally small mass $m$, we will elaborate on the near-boundary behavior for the bulk fields later on in section \ref{holo_renor}.

Importantly, the bulk path integral \eqref{Z_AdS1} can be equivalently expressed in terms of gauge-fixed configurations (once the gauge transformation parameter is included)
\begin{align}
    Z_{\rm AdS} & = \int [D C_\mu] [D\xi] [D\theta] e^{i S_{\rm bulk}[C_\mu, \theta] + i S_{\rm ct}} = \int [D\xi] e^{i S_{\rm bulk}[C_\mu[B_\mu, \vartheta], \, \theta[B_\mu,\vartheta]]\big|_{\rm p.o.s} + i S_{\rm ct}} \nonumber \\
    & \approx \int [D \phi] e^{i S_{\rm bulk}[C_\mu[B_\mu, \vartheta], \, \theta[B_\mu,\vartheta]]\big|_{\rm p.o.s} + i S_{\rm ct}} , \label{Z_AdS2}
\end{align}
where the Jacobian determinant, arising from going from $(C_M^\prime,\theta^\prime)$ to $(C_r= - C_v/(r^2f(r)), C_\mu, \theta)$, is neglected (hence, the $\approx$). In the right-handed side of the first line in \eqref{Z_AdS2}, the heavy modes dual to $C_\mu, \theta$ have been integrated out in the saddle point approximation, yielding the partially on-shell (p.o.s) action $S_{\rm bulk}[C_\mu[B_\mu, \vartheta], \theta[B_\mu,\vartheta]]\big|_{\rm p.o.s}$, which is eventually identified as the desired effective action for the boundary theory
\begin{align}
    S_{eff} = S_{\rm bulk}[C_\mu[B_\mu, \vartheta], \, \theta[B_\mu,\vartheta]]\big|_{\rm p.o.s} + S_{\rm ct}.
\end{align}
Here, in the partially on-shell action, we shall not impose the constraint equation, which corresponds to the ``conservation law'' for the boundary current. More details will be provided in section \ref{bulk_dynamics}.

The analysis above can be slightly modified in order to be compatible with a different gauge choice, \textit{e.g.}, $C_r = 0$. While some details might change, the main conclusion does not.

\subsection{Bulk dynamics: variational problem revisited} \label{bulk_dynamics}

In the saddle point approximation, the bulk dynamics reduces to solving the classical equations of motion for the bulk fields $C_M$ and $\theta$. However, we shall do this in a partially on-shell approach so that the low energy mode will be kept as dynamical in the effective action. It turns out that this prescription becomes more natural by reconsidering the bulk variational problem based on gauge-fixed field configurations.

First of all, it is direct to carry out such a procedure based on the field configuration $(C_M^\prime, \theta^\prime)$, which can be varied freely
\begin{align}
    C_M^\prime \to C_M^\prime + \delta C_M^\prime, \qquad \theta^\prime \to \theta^\prime + \delta \theta^\prime.
\end{align}
Therefore, the variation of the bulk action reads
\begin{align}
    \delta S_{\rm bulk} = & \int d^5x \sqrt{-g} \left[\nabla_M  F^{\prime MN} \delta C_N^\prime - \delta C_M^\prime m^2(C^{\prime M} - \nabla^M \theta^\prime) - m^2 \nabla_M  (C^{\prime M} - \nabla^M \theta^\prime) \delta \theta^\prime \right] \nonumber \\
   & + S_{\rm bdy} \nonumber \\
  = & \int d^5x \sqrt{-g} \left[\nabla_M  F^{MN} \delta C_N^\prime - \delta C_M^\prime m^2(C^{M} - \nabla^M \theta) - m^2 \nabla_M  (C^{M} - \nabla^M \theta) \delta \theta^\prime \right] \nonumber \\
   & + S_{\rm bdy} ,
\end{align}
where $S_{\rm bdy}$ is a potential boundary term which is irrelevant for the subsequent discussion.  Now, we are ready to  express $\delta S_{\rm bulk}$ in terms of a gauge-fixed configuration. To this end, via \eqref{gauge_transformation1}, we have
\begin{align}
    \delta C_r^\prime = - \frac{\delta C_v}{r^2f(r)} - \nabla_r \delta \xi, \qquad \delta C_\mu^\prime = \delta C_\mu - \nabla_\mu \delta \xi, \qquad \delta\theta^\prime = \delta \theta - \delta \xi .
\end{align}
Thus, from $\delta S_{\rm bulk}$, we obtain the dynamical equations of motion (EOMs)
\begin{align}
    &\delta C_v \neq 0 \Rightarrow \nabla_M F^{M v} - m^2 (C^v - \nabla^v \theta)  - \frac{1}{r^2f(r)} \left[ \nabla_M F^{M r} - m^2 (C^r - \nabla^r \theta) \right] =0, \nonumber \\
    &\delta C_i \neq 0 \Rightarrow \nabla_M F^{M i} - m^2 (C^i - \nabla^i \theta)=0, \nonumber \\
    &\delta \theta \neq 0 \Rightarrow \nabla_M (\nabla^M \theta - C^M) =0,  \label{dynamical_eom}
\end{align}
and the contracted Bianchi identity
\begin{align}
    \delta \xi \neq 0 \Rightarrow \nabla_M \nabla_N F^{MN} =0 . \label{Bianchi}
\end{align}
Lastly, we also have a potential boundary term due to variation of the gauge parameter, $\delta \xi$, which is given by
\begin{align}
    \delta S_{\rm bulk} \rightarrow \int d^4x \sqrt{-\gamma} n_N \left[ - \nabla_M F^{MN} + m^2 (C^N - \nabla^N \theta) \right] \delta \xi \big|_{\rm bdy}.
\end{align}
The latter gives the constraint equation
\begin{align}
    \delta \xi\big|_{\rm bdy} \neq 0 \Rightarrow \mathbb{E}^r \equiv\nabla_M F^{Mr} - m^2 (C^r - \nabla^r \theta)\big|_{\rm bdy} =0.  \label{constraint}
\end{align}

Using the radial gauge
\begin{equation}
C_r= - \frac{C_v}{r^2f(r)}\,,    
\end{equation}
the dynamical equations \eqref{dynamical_eom} fully determine the profiles of bulk fields $C_M$ and $\theta$ in the bulk spacetime. Indeed, after solving the dynamical equations \eqref{dynamical_eom}, the value of the constraint $\mathbb{E}^r$ will become known at any spacetime point. Moreover, the Bianchi identity \eqref{Bianchi} allows to obtain a compact expression for $\mathbb{E}^r$ at any radial slice.

Then, explicitly, the dynamical equations \eqref{dynamical_eom} are
\begin{align}
    0= &\partial_r(r^3 \partial_r C_v) + \left[ \frac{2r}{f(r)} \partial_r + \frac{1}{f(r)} - \frac{rf^\prime(r)}{f^2(r)} \right] \partial_v C_v + \frac{\partial_v^2 C_v}{rf^2(r)} + \frac{1}{rf(r)} (\partial_k^2 C_v - \partial_v \partial_k C_k) \nonumber \\
    & - \frac{m^2 r}{ f(r)} (C_v - \partial_v \theta), \nonumber \\
   0 = & \partial_r [r^3f(r) \partial_r C_i] + 2 r \partial_r \partial_v C_i + \partial_v C_i + \frac{\partial_v \partial_i C_v}{rf(r)} 
   + r^{-1} \partial_k (\partial_k C_i - \partial_i C_k) - m^2 r(C_i - \partial_i \theta), \nonumber \\
   0 = & \partial_r [r^5f(r) \partial_r \theta] + 2r^3 \partial_r \partial_v \theta + 3r^2 \partial_v \theta  + r\partial_k (\partial_k \theta - C_k) + \frac{r}{f(r)} \partial_v C_v. \label{eom_CvCitheta}
\end{align}
These EOMs will be solved on the radial contour shown in Figure \ref{holographic_SK_contour}.

\subsection{Holographic renormalization} \label{holo_renor}

We turn to the near-boundary behavior for the bulk fields. Remarkably, the asymptotic behavior for the bulk fields depends on the value of bulk mass $m$. The present work will be limited to the case of a small $m$. Then, within this assumption, Eq.~\eqref{eom_CvCitheta} will be solved perturbatively
\begin{align}
    C_\mu = C_\mu^{(0)} + m^2 C_\mu^{(2)} + \cdots, \qquad \qquad \qquad \theta = \theta^{(0)} + m^2 \theta^{(2)} + \cdots. \label{Cmu_theta_m-expansion}
\end{align}
Then, near the AdS boundary, we have
\begin{align}
  &  C_\mu^{(0)}(r\to \infty,x^\alpha) = \underline{ B_\mu } + \frac{\partial_v B_\mu}{r} - \frac{\log r }{2r^2} \partial^\nu (\partial_\mu B_\nu -  \partial_\nu B_\mu) + \frac{\tilde J_\mu^{(0)}}{r^2} + \cdots, \nonumber \\
   & C_\mu^{(2)}(r \to \infty, x^\alpha) =  \frac{1}{2} (B_\mu - \partial_\mu \vartheta) \log r + \underline{ 0 } + \frac{\log r}{2r} \partial_v (B_\mu - \partial_\mu \vartheta )  + \frac{\log r}{8r^2} \delta B_\mu^{\rm L} + \frac{\tilde J_\mu^{(2)}}{r^2} + \cdots, \nonumber \\
  &  \theta^{(0)}(r\to \infty,x^\alpha) = \underline{ \vartheta } + \frac{\theta_1}{r} + \frac{\theta_2}{r^2} + \frac{\theta_3}{r^3} + \frac{\log r}{r^4}\theta_L + \frac{\tilde O^{(0)}}{r^4} + \cdots, \label{Cmu_theta_bdy}
\end{align}  
where
\begin{align}
    & \delta B_\mu^{\rm L} = -4 \tilde J_\mu^{(0)} - \partial_\mu \partial_\nu B^\nu + 4 \partial_v^2 B_\mu + \partial_\mu ({\vec \partial}^{\, 2} -3 \partial_v^2) \vartheta, \nonumber \\
    & \theta_1 = \partial_v \vartheta, \nonumber \\
    & \theta_2 = \frac{1}{4} \left(\partial_v^2 + {\vec\partial}^{\,2} \right) \vartheta - \frac{1}{4} \partial_\mu B^\mu , \nonumber \\
    & \theta_3 = \frac{1}{12} \left( 3 {\vec\partial}^{\,2} - \partial_v^2 \right) \partial_v \vartheta - \frac{1}{4} \partial_v \partial_\mu B^\mu, \nonumber \\
    & \theta_L = - \frac{1}{4} \partial_\mu \tilde J^{\mu (0)}  + \frac{1}{16} \left( {\vec \partial}^{\,2} -3 \partial_v^2 \right) \partial_\mu B^\mu  + \frac{1}{16} \left( {\vec \partial}^{\,2} - \partial_v^2 \right)^2 \vartheta.
\end{align}
Since we will truncate the boundary action up to $\mathcal{O}(m^2)$, the knowledge of $\theta^{(2)}$ is irrelevant. The underlined terms in \eqref{Cmu_theta_bdy} indicate the free boundary data which will be fixed by AdS boundary conditions. We will elaborate on this point in section \ref{perturb_scheme}.

We proceed by explaining the derivation of the counter-term action $S_{\rm ct}$, by applying the standard procedure of holographic renormalization \cite{Bianchi:2001kw,Skenderis:2002wp}. Once the dynamical equations \eqref{dynamical_eom} are solved, it is direct to evaluate the partially on-shell bulk action. For a quadratic action like \eqref{Stueckelberg_action}, it is possible to reduce it into a surface term via integration by parts. It turns out that
\begin{align}
    S_{\rm bulk}= & - \frac{1}{2} \int d^4x \sqrt{-\gamma} n_M \left[ C_N F^{MN} + m^2 \theta (\nabla^M \theta - C^M) \right]\bigg|_{r=\infty_2}^{r=\infty_1} \nonumber \\
    = & - \frac{1}{2} \int d^4x \sqrt{-g} \left[ C_\mu F^{r \mu} + m^2 \theta (\nabla^r \theta - C^r) \right]\bigg|_{r=\infty_2}^{r=\infty_1}, \nonumber \\
    = & - \frac{1}{2} \int d^4x \left\{ - r^3 C_v \partial_r C_v - \frac{r}{f(r)} C_v \partial_v C_v + r^3f(r) C_k \partial_r C_k + r C_k \partial_v C_k \right. \nonumber \\ 
   &\qquad \qquad \qquad \left. + m^2 r^5f(r) \theta \partial_r \theta + m^2 r^3 \theta \partial_v \theta \right\}. \label{S_surface}
\end{align}
Here, we utilized the dynamical equations \eqref{dynamical_eom} and the radial gauge choice, but have not assumed the constraint equation \eqref{constraint}. Using the near-boundary behavior \eqref{Cmu_theta_bdy}, it is straightforward to obtain the divergent part of the bulk action \eqref{Stueckelberg_action} near to the conformal boundary $r=\infty$. The result is given by
\begin{align}
    S_{\rm div} = &  \int d^4x \frac{\log r_c}{4}  (\partial_\mu B_\nu - \partial_\nu B_\mu) (\partial^\mu B^\nu - \partial^\nu B^\mu) \nonumber \\
    & + m^2 \int d^4x \left\{ - \frac{1}{4} r_c^2 (B_\mu - \partial_\mu \vartheta) (B^\mu - \partial^\mu \vartheta) - \frac{1}{8} (\log^2 r_c - \log r_c) (\partial_\mu B_\nu - \partial_\nu B_\mu)  \right. \nonumber \\
    & \qquad \qquad \qquad \left. \times (\partial^\mu B^\nu - \partial^\nu B^\mu) + \frac{1}{8} \log r_c\, [\partial_\mu (B^\mu - \partial^\mu \vartheta) ]^2 \right\}, \label{Sdiv}
\end{align}
where $r_c$ is a UV cutoff. In a minimal subtraction scheme, the counter-term action $S_{\rm ct}$ may be taken as
\begin{align}
    S_{\rm ct} = & \int d^4x  \sqrt{-\gamma} \left\{ - \frac{\log r} {8} (2-3m^2 \log r) F_{\mu\nu} F^{\mu\nu} + \frac{1}{4} m^2 (C_\mu - \partial_\mu \theta) (C^\mu - \partial^\mu \theta) \right. \nonumber \\
    & \qquad \qquad \qquad  \left. - \frac{\log r}{8} m^2 \left[ \partial_\mu (C^\mu - \partial^\mu \theta) \right]^2  \right\}\bigg|_{r= \infty}. \label{Sct}
\end{align}
 In the counter-term action \eqref{Sct}, while the first piece and the third piece are written down based on minimal subtraction scheme, the second piece will unavoidably bring in {\rm finite} contributions when $r \to \infty$. Here, we briefly compare our counter-term action \eqref{Sct} with that proposed in \cite{Jimenez-Alba:2014iia,Jimenez-Alba:2015awa}
\begin{align}
    S_{\rm ct}^{\rm JA} = \int d^4x \sqrt{-\gamma} \left\{ \frac{\Delta}{2} (C_\mu - \partial_\mu \theta) (C^\mu - \partial^\mu \theta) - \frac{1}{4(\Delta +2)} [\partial_\mu (C^\mu - \partial^\mu \theta)]^2 + \frac{1}{8\Delta} F^2 \right\} \label{Sct_JA}.
\end{align}
While \eqref{Sct_JA} does cancel most of the divergences in \eqref{Sdiv}, it becomes ill-defined as $m\to 0$. Eventually, the boundary EFT action is given by
\begin{align}
    S_{\rm eff} = \int d^4x & \left[ B_\mu(x) \left( \tilde J^{\mu (0)}(x) + \frac{1}{2} m^2 \tilde J^{\mu (0)}(x) + m^2 \tilde J^{\mu (2)}(x) \right) + 2 m^2 \vartheta(x) \tilde O^{(0)}(x)  \right. \nonumber \\
    & \quad \left. - \frac{3}{8} m^2 \tilde J^{\mu (0)} \partial_\mu \vartheta + \cdots \right] \bigg|_{2}^1\,\, ,  \label{Seff_formal}
\end{align}
where $\cdots$ denotes contact terms beyond order $\mathcal{O}(\partial_\mu^2)$. As evident, Eq.~\eqref{Seff_formal} has a smooth $m \rightarrow 0$ limit.

\subsection{Boundary conditions and perturbative scheme} \label{perturb_scheme}

We will impose AdS boundary conditions and fix the underlined terms in \eqref{Cmu_theta_bdy}. However, in order to fully fix $C_v$, we shall additionally impose a vanishing condition for $C_v$ at the horizon \cite{Glorioso:2018mmw}
\begin{align}
    C_v(r=r_h,x^\alpha) =0\,.  \label{Cv=0}
\end{align}
This condition corresponds to the chemical shift symmetry in \eqref{chemical_shift}, which becomes more obvious if we examine the residual gauge symmetry. Indeed, with the radial gauge-fixing $C_r =- C_v/(r^2f(r))$, we still have a residual gauge symmetry for the configuration $(C_M, \theta)$
\begin{align}
    C_\mu \to C_\mu + \partial_\mu \Lambda(x^\alpha), \qquad  \theta \to \theta + \Lambda(x^\alpha), \label{residual_gauge_symmetry}
\end{align}
where $\Lambda(x^\alpha)$ is to distinguish from the gauge transformation parameterized by $\xi$ that has brought $(C_M^\prime, \theta^\prime)$ to $(C_M, \theta)$. Now, the condition \eqref{Cv=0} requires $\Lambda$ to be time-independent, $\Lambda(x^\alpha) \to \Lambda(\vec x)$. Eventually, the boundary version of \eqref{residual_gauge_symmetry} is nothing else but the chemical shift symmetry \eqref{chemical_shift}.

Note however that preserving chemical shift symmetry actually allows for a more general condition at the horizon, which can be generally written as
$C_v(r_h,x^\alpha) =F(x^\alpha)$  
for any fixed function $F$ of the boundary coordinates $x^\alpha$. It is tempting to relate such a redundancy in the horizon condition to the freedom of choosing a specific hydrodynamic frame. In other words, we expect that the choice of the $F$ function above would entirely fix the zero modes of the bulk differential operators derived within the fluid-gravity correspondence, which indeed encode the hydrodynamic frame transformations \cite{Hoult:2021gnb}.\footnote{See also \cite{Donos:2017ihe, Bu:2015ame, Donos:2022uea} for the appearance of these zero modes in explicit holographic constructions of hydrodynamic modes and for discussions on hydrodynamic frames in related holographic settings.} We leave such an exploration as a future task.

In all our discussions, we assume that the bulk mass $m$ is small. This is equivalent to assuming that the explicit breaking term in the dual field theory is small compared to the characteristic microscopic scale of the system. More specifically, we will work within the approximation of $m/T \ll 1$, where $T$ is the temperature of the dual field theory.
Within this assumption, \eqref{eom_CvCitheta} can be solved perturbatively, \textit{cfr}. Eq.~\eqref{Cmu_theta_m-expansion}. At each order in the $m$-expansion of \eqref{Cmu_theta_m-expansion}, we perform a hydrodynamic expansion
\begin{align}
    &C_\mu^{(0)} =\varepsilon \,C_\mu^{(0)(1)} + \varepsilon^2\, C_\mu^{(0)(2)} + \cdots, \qquad \qquad C_\mu^{(2)} = \varepsilon\, C_\mu^{(2)(1)} + \varepsilon^2\, C_\mu^{(2)(2)} + \cdots, \nonumber \\
    &\theta^{(0)} = \theta^{(0)(0)} + \varepsilon\, \theta^{(0)(1)} + \varepsilon^2 \theta^{(0)(2)} + \cdots, \label{derivative_expansion}
\end{align}
where the bookkeeping parameter $\varepsilon \sim \partial_\mu$ helps to count number of boundary derivatives. In Fourier space, $\varepsilon$ has to be identified with the frequency $\omega$ and the wave-vector $k$. Here, the derivative expansion in \eqref{derivative_expansion} is motivated by the fact that $B_\mu = A_\mu + \partial_\mu \phi $ is first order in boundary derivatives, while $\vartheta = \theta_b + \phi$ is zeroth order in boundary derivative.
Notice that, at this point, we are performing a double and independent expansion in the parameter $m$ and the spacetime derivatives $\partial_\mu$, hence the double indices.

Consequently, at each order in the double expansion, $C_\mu^{(n)(l)}$ and $\theta^{(n)(l)}$ satisfy decoupled ordinary differential equations (ODEs)
\begin{align}
   \partial_r \left[ r^3 \partial_r C_v^{(n)(l)} \right] = j_v^{(n)(l)}, \quad
   \partial_r \left[ r^3 f(r) \partial_r C_i^{(n)(l)} \right] = j_i^{(n)(l)}, \quad
   \partial_r \left[ r^5f(r) \partial_r \theta^{(n)(l)} \right] = j_\theta^{(n)(l)}, \label{ODEs}
\end{align}
where the sources are built up from lower order solutions. For our purpose, we record the first few source terms
\begin{align}
    &j_\theta^{(0)(0)}=  j_v^{(0)(1)} = j_i^{(0)(1)} = 0, \nonumber \\
    &j_\theta^{(0)(1)} = - 2r^3 \partial_r \partial_v \theta^{(0)(0)} - 3r^2 \partial_v \theta^{(0)(0)}, \nonumber \\
    & j_v^{(0)(2)} = -  \left[ \frac{2r}{f(r)} \partial_r + \frac{1}{f(r)} - \frac{rf^\prime(r)}{f^2(r)} \right] \partial_v C_v^{(0)(1)}, \nonumber \\
    & j_i^{(0)(2)} = - 2r \partial_r \partial_v C_i^{(0)(1)} - \partial_v C_i^{(0)(1)}, \nonumber \\
    &j_v^{(2)(1)} = \frac{r}{f(r)} \left[  C_v^{(0)(1)} - \partial_v \theta^{(0)(0)} \right], \nonumber\\
    &j_i^{(2)(1)} = r \left[ C_i^{(0)(1)} - \partial_i \theta^{(0)(0)} \right], \nonumber \\
    &j_v^{(2)(2)} = -  \left[ \frac{2r}{f(r)} \partial_r + \frac{1}{f(r)} - \frac{rf^\prime(r)}{f^2(r)} \right] \partial_v C_v^{(2)(1)} + \frac{r}{f(r)} \left[  C_v^{(0)(2)} - \partial_v \theta^{(0)(1)} \right], \nonumber \\
    &j_i^{(2)(2)} = - 2r \partial_r \partial_v C_i^{(2)(1)} - \partial_v C_i^{(2)(1)} + r \left[ C_i^{(0)(2)} - \partial_i \theta^{(0)(1)} \right], \nonumber \\
    &j_\theta^{(0)(2)} = -2r^3 \partial_r \partial_v \theta^{(0)(1)} - 3r^2 \partial_v \theta^{(0)(1)} - r \partial_k \left(\partial_k \theta^{(0)(0)} - C_k^{(0)(1)} \right) - \frac{r}{f(r)} \partial_v C_v^{(0)(1)}. \label{source_terms}
\end{align}
The double expansion has been truncated as follows. First, since $\theta$ is always multiplied by a $m^2$ factor, we will keep its lowest order in the $m$-expansion. Then, we cover the derivative terms in $\theta^{(0)}$ up to second order. Second, ignoring the backreaction of $\theta$ on the bulk gauge field, we will construct $C_\mu^{(0)}$ up to second order in boundary derivative expansion. Finally, the backreaction effect of $\theta$ on the dynamics of the bulk gauge field is important, and will not be neglected. For this purpose, we track the leading order $m^2$-corrections $C_\mu^{(2)}$ up to second order in boundary derivative expansion.

The truncation scheme specified above guarantees that we will obtain the structure of the EFT Lagrangian \eqref{L1}, along with the specific values for the coefficients. Since the perturbative computations are rather cumbersome and not very illuminating, we will defer all the details and results to Appendix \ref{app1}.

\subsection{Summary of results}\label{sec:hol_results}

With the perturbative solutions presented in appendix \ref{app1}, it is straightforward to evaluate \eqref{Seff_formal}. Limited to the truncated order, we do obtain the desired EFT structure as in Eq.~\eqref{L1} in which the various coefficients are given by
\begin{align}
    &g_0 = 2r_h^2 + \frac{1}{2} m^2 r_h^2 \left[ 2\log(2r_h^2) +1 \right],\quad g_1 =0,\quad f_0 = \frac{1}{2} m^2 r_h^2,\quad f_1 = -r_h - \frac{1}{2} m^2 r_h \log(2r_h^2),  \nonumber \\
    &h_0=0,\quad h_1 =0,\quad \Gamma_0 = 0,\quad \tilde \Gamma = -m^2 r_h^3,\quad \Gamma_1 =  \frac{1}{2}m^2 r_h^2 (1-\log 2),\quad \Gamma_2 =- \frac{1}{2} m^2 r_h^2, \nonumber \\
    &c_0 =0,\quad c_1 = \frac{1}{2} m^2 r_h^2 \log(2r_h^2),\quad c_2 =  \frac{1}{2} m^2 r_h^2,\quad d_0=0,\quad d_1 =  -\frac{1}{2} m^2 r_h^2 \log(2r_h^2), \nonumber \\
    &d_2 = -\frac{1}{2} m^2 r_h^2,\quad w_0 = 0,\quad w_1 = 0,\quad w_2 = \frac{2ir_h^2}{\pi} + m^2 \frac{ir_h^2}{\pi} \log(2r_h^2),\quad w_3 = 2m^2 \frac{ir_h^4}{\pi}, \nonumber \\
    &w_4 = m^2 \frac{i r_h^2 \log 2}{\pi},\quad w_5 = -m^2 \frac{i r_h^2}{24\pi} \left( 5\pi^2 +24\log2 -12 \log^22 \right),\quad w_6=0, \nonumber \\
    &w_7 = m^2\frac{ir_h^2}{4\pi} (2 \log2 - \log^2 2),\quad w_8 = m^2 \frac{ir_h^2}{\pi} \log r_h.  \label{holo_SK_coeffs}
\end{align}
We have checked that the holographic results \eqref{holo_SK_coeffs} indeed satisfy all the constraints summarized in section \ref{sec1}. Notice that the AdS length scale is set to $L_{AdS}=1$; restoring it will make the arguments in the logarithms dimensionless. Before continuing, let us discuss a few important points related to the various transport coefficients above.

\begin{itemize}
    \item In the absence of explicit breaking, $f_1$ and $g_0$ correspond respectively to the electric conductivity and the charge susceptibility, with their ratio giving the charge diffusion constant.
    \item The vanishing of the $g_1$ coefficient, $g_1=0$, which was also observed in \cite{Glorioso:2018mmw,deBoer:2018qqm,Bu:2020jfo}, is related to the hydrodynamic frame that the holographic models naturally choose. Indeed, in the absence of explicit breaking, $g_1$ can be consistently set to zero by an appropriate redefinition of the dynamical field, at the cost of having an additional higher-order terms. In other words, its value is not frame independent. On the contrary, in the case of spontaneous or pseudo-spontaneous breaking, chemical shifts are not anymore a symmetry of the system and the corresponding horizon boundary condition is lost. As a matter of fact, for superfluids, $g_1$ gives the Goldstone diffusivity, which is a ``measurable'' transport coefficient (see for example \cite{Delacretaz:2021qqu}). 
    \item In the EFT language, $\Gamma_0=0$ corresponds to the vanishing of the so-called pinning frequency. This is expected since there is no spontaneous symmetry breaking in our model and therefore no Goldstone mode to be pinned.
    \item $\tilde \Gamma$ determines the relaxation rate of charge and will play a major role in the analysis of section \ref{sec3}.

    \item In section \ref{sec:aj_coefficients}, we will comment on the relationship between some of other coefficients we found, particularly $\Gamma_2, c_1, \Gamma_1$ and the $m^2$-part of $g_0$ (or equivalently $c_2$, $d_2$, $d_1$), and some novel coefficients proposed in \cite{Armas:2021vku} (see also \cite{Delacretaz:2021qqu}).
\end{itemize}

\section{Lessons from effective field theory and holography} \label{sec3}

After explaining in detail both the EFT and holographic sides, we are now in a position to bring all the results together and present them in a coherent form. The purpose of this section is twofold: present a derivation of the non-conservation Ward identity using EFT and predict the dispersion of the relaxed charge diffusion mode.

\subsection{The broken U(1) Ward identity}\label{subsec:ward}

We define the non-conserved $U(1)$ currents as the functional derivative of EFT action with respect to the external gauge field
\begin{align}
    J_r^\mu \equiv \frac{\delta S_{\rm eff}}{\delta A_{a\mu}}, \qquad \qquad  J_a^\mu \equiv \frac{\delta S_{\rm eff}}{\delta A_{r\mu}}.
\end{align}
Meanwhile, the expectation values of the scalar operator (dual to the bulk scalar field $\theta$) are
\begin{align}
    O_r \equiv \frac{\delta S_{\rm eff}}{\delta \theta_{ba}}, \qquad \qquad  O_a \equiv \frac{\delta S_{\rm eff}}{\delta \theta_{br}}.
\end{align}
For simplicity, we will take $\mathcal L_1$ to illustrate the derivation. Then, the results are
\begin{align}
    &J_r^0 = g_0 B_{r0} + g_1 \partial_0 B_{r0} + d_1 \partial_0 \vartheta_r + w_0 B_{a0} - w_7 \partial_0 \vartheta_a, \nonumber \\
    &J_r^i = f_0 B_{ri} + f_1 \partial_0 B_{ri} + d_2 \partial_i \vartheta_r + w_2 B_{ai} - w_8 \partial_i \vartheta_a, \nonumber \\
    &J_a^0 = g_0 B_{a0} - g_1 \partial_0 B_{a0} - c_1 \partial_0 \vartheta_a, \nonumber \\
    &J_a^i = f_0 B_{ai} - f_1 \partial_0 B_{ai} - c_2 \partial_i \vartheta_a, \nonumber \\
    &O_r = \tilde \Gamma \partial_0 \vartheta_r + \Gamma_1 \partial_0^2 \vartheta_r + \Gamma_2 \partial_k^2 \vartheta_r + c_1 \partial_0 B_{r0} + c_2 \partial_i B_{ri} + w_3 \vartheta_a + w_4 \partial_k^2 \vartheta_a \nonumber \\
    & \qquad + w_5 \partial_0^2 \vartheta_a + w_7 \partial_0 B_{a0} + w_8 \partial_i B_{ai}, \nonumber \\
    & O_a = -\tilde \Gamma \partial_0 \vartheta_a + \Gamma_1 \partial_0^2 \vartheta_a + \Gamma_2 \partial_k^2 \vartheta_a - d_1 \partial_0 B_{a0} - d_2 \partial_i B_{ai}.  \label{Jmu_O}
\end{align}
The equations of motion for dynamical fields $\phi_a$ and $\phi_r$ are simply the Ward identities for the currents defined above
\begin{align}
    \frac{\delta S_{\rm eff}}{\delta \phi_a} =0 \Rightarrow \partial_\mu J_r^\mu = O_r, \qquad \qquad \frac{\delta S_{\rm eff}}{\delta \phi_r} =0 \Rightarrow \partial_\mu J_a^\mu =  O_a.
\end{align}

The physical current $J_r^\mu$ and scalar operator's expectation value $O_r$ can be split into a mean-field part and a stochastic part
\begin{align}
    J_r^\mu = J_{\rm mf}^\mu + J_{\rm st}^\mu, \qquad \qquad  O_r = O_{\rm mf} + O_{\rm st},
\end{align}
corresponding to the hydrodynamic current and the noise respectively. Here, we can perform a Legendre transformation to turn the variable $\phi_a$ into $\zeta$, which now obeys Gaussian distribution (see \cite{Crossley:2015evo,Bu:2020jfo} for more details). Then, the Ward identity for the physical current can be rewritten into a Langevin-type equation
\begin{align}
    \partial_\mu J_{\rm mf}^\mu = O_{\rm mf} + \zeta,
\end{align}
where $\zeta$ is the stochastic force.

Therefore, when the noise term is ignored, the Ward identity becomes
\begin{align}
    \partial_\mu J_{\rm mf}^\mu = O_{\rm mf},
\end{align}
where
\begin{align}
     &J_{\rm mf}^0 = g_0 B_{r0} + g_1 \partial_0 B_{r0} + d_1 \partial_0 \vartheta_r, \nonumber \\
    &J_{\rm mf}^i = f_0 B_{ri} + f_1 \partial_0 B_{ri} + d_2 \partial_i \vartheta_r, \nonumber \\
    &O_{\rm mf} = \tilde \Gamma \partial_0 \vartheta_r + \Gamma_1 \partial_0^2 \vartheta_r + \Gamma_2 \partial_k^2 \vartheta_r + c_1 \partial_0 B_{r0} + c_2 \partial_i B_{ri}.
\end{align}

As an approximation, we focus on the leading order term $\tilde \Gamma \partial_0 \vartheta_r$ in $O_{\rm mf}$:
\begin{align}
   \tilde  \Gamma \partial_0 \vartheta_r \to \tilde \Gamma \partial_0 \phi_r \approx \frac{\tilde \Gamma}{g_0 + d_1} J_{\rm mf}^0\,.
\end{align}
Then, the Ward identity simplifies to
\begin{align}
    \partial_\mu J_{\rm mf}^\mu = - \Gamma J_{\rm mf}^0 +\cdots\,, \label{ward_id_derivation}
\end{align}
where the relaxation rate of the $U(1)$ current is
\begin{align}\label{ops}
   \Gamma = - \frac{\tilde \Gamma}{g_0 + d_1} \approx  \frac{1}{2} m^2 r_h = \frac{1}{2} m^2 \pi T,
\end{align}
which will be further confirmed by numerical study in section \ref{numeric_study}. We thus see how \eqref{eq2} emerges in our field theoretic SK construction in a similar way as discussed in section \ref{sec:u1intro}.

\subsection{Novel coefficients} \label{sec:aj_coefficients}

In this section, we would like to clarify the relationship between some coefficients in our analysis and some new coefficients proposed in \cite{Armas:2021vku} (\textit{e.g.}, $\lambda,\bar f_s$ therein).

First, let us discuss the dissipative nature of the various coefficients in the EFT Lagrangian \eqref{L1}. Inspired by the old Martin-Siggia-Rose formalism \cite{PhysRevA.8.423}, ref.~\cite{Kapustin:2022iih} recently showed how to separately construct the non-dissipative and dissipative KMS-invariant terms in the effective Lagrangian
\begin{align}
    \mathcal L = \mathcal L^{nd} + \mathcal L^d.
\end{align}
The non-dissipative part $\mathcal L^{nd}$ can be derived from a real Lagrangian $\Omega[\varphi_r]$ by setting
\begin{equation}
    \mathcal L^{nd} = \varphi_a \frac{\partial\Omega}{\partial\varphi_r},
\end{equation}
where we use $\varphi_r$ and $\varphi_a$ to denote all $r$-type and $a$-type fields (including sources), respectively. The dissipative part $\mathcal L^d$ is derived from a suitable, quadratic in $\varphi_a$, quantity $X$ as
\begin{equation}
    \mathcal L^d = \frac{1}{2}\left[ X(\varphi_r,\varphi_a) + X_{KMS}(\varphi_r,\varphi_a) -X(\varphi_r,0) - X_{KMS}(\varphi_r,0) \right],
\end{equation}
where $X_{KMS}$ is what we obtain after applying the KMS transformation \eqref{classical_KMS_transform} on $X$.

In our case, it is useful to define the gauge invariant source combination
\begin{align}\label{eq:xi_def}
    \Xi_{s\,\mu} \equiv \partial_\mu \mathcal{\theta}_{b\,s} - A_{s\,\mu} = \partial_\mu \vartheta_s - B_{s\mu}.
\end{align}
Using \eqref{eq:xi_def}, we can equivalently build the effective action in terms of $B_{s\,\mu}$, $\Xi_{s\,\mu}$, and $\vartheta_{s}$ without derivatives acting on the latter. In fact, in order to construct the relevant $ra$-terms of $\mathcal{L}_1$ in \eqref{L1}, we can use the seed functions $\Omega$ and $X$
\begin{align}
    \Omega &= \left(g_{0}+\Gamma_1+2c_1\right) B_{r0}^2 +\left(\Gamma_1+c_1\right) B_{r0}\Xi_{r0} +\Gamma_1 \Xi_{r0}^2 +\Gamma_2 \Xi_{ri}^2\,, \nonumber\\
    X &= \frac{1}{i\beta} \left(g_{1} B_{a0}^2 +f_{1} B_{ai}^2 +\tilde{\Gamma} \vartheta_a^2 \right)\,. \label{Omega_X}
\end{align}

Eq. \eqref{Omega_X} clearly shows that the coefficients $g_{0},c_1,\Gamma_1$ and $\Gamma_2$ are non-dissipative, while $g_{1},f_{1}$ and $\tilde{\Gamma}$ are dissipative. Then, following the procedure outlined above, we obtain the alternative form of $\mathcal{L}_1$ 
\begin{align}
\mathcal{L}_1^{(ra)}&=\left(g_{0}+\Gamma_1+2c_1\right) B_{a0}B_{r0}+g_{1} B_{a0}\partial_0 B_{r0} + \left(\Gamma_1+c_1\right)   (B_{a0}\Xi_{r0} + \Xi_{a0}B_{r0}) \nonumber \\
    & + f_1 B_{ai}\partial_0 B_{ri} +\tilde{\Gamma} \vartheta_a (B_{r0}+\Xi_{r0}) +\Gamma_1 \Xi_{a0}\Xi_{r0} -\Gamma_2 \Xi_{ai}\Xi_{ri}\,, \label{eq:app_l0}
\end{align}
which automatically satisfies the chemical shift symmetry, the KMS conditions and the Onsager relations, \eqref{chemical_shift} and \eqref{Onsager_relation}. We recall that $g_0,g_1,f_1$ have contributions of order $\mathcal{O}(m^0)$ along with $\mathcal{O}(m^2)$ corrections, whereas $\tilde \Gamma,\Gamma_1,\Gamma_2,c_1$ are of order $\mathcal{O}(m^2)$ and they appear only after explicitly breaking the global $U(1)$ symmetry.

From \eqref{eq:app_l0}, one can observe that $\Gamma_1$ and $\Gamma_2$ are contact-type coefficients. These terms are not shown in \cite{Armas:2021vku} because they are subleading in their derivative counting. On the other hand, here, the coefficient labelled as $\bar{f}_s$ in \cite{Armas:2021vku} does not appear, because we are in the normal phase and chemical shift symmetry implies that it vanishes identically.\footnote{We would like to thank Akash Jain and Jay Armas for discussions on this point.} The relation between the other coefficients found in our setup and the one denoted by $\lambda$ in \cite{Armas:2021vku} can be understood as follows.

Turning the sources $\Xi_{s\,\mu}$ off, the constitutive relations at leading order take the form
\begin{align}
     &J_{\rm mf}^0 = (g_0+c_1) B_{r0} + g_1 \partial_0 B_{r0}, \nonumber \\
    &J_{\rm mf}^i = f_1 (\partial_i B_{r0} + F_{0i}), \nonumber \\
    &O_{\rm mf} = \tilde{\Gamma} B_{r0} -(\Gamma_1+c_1)\partial_0 B_{r0}\,.
\end{align}
Using a frame redefinition, we could define the chemical potential $\mu$ by
\begin{align}\label{eq:frame}
    g^{(0)}_0 \mu\equiv (g_0+c_1) B_{r0}\,,
\end{align}
where $g_0^{(0)}$ represents the $m^0$-part of $g_0$. This leads to
\begin{align}\label{eq:frame_cr}
    &J_{\rm mf}^0 = g^{(0)}_0 \mu +\dots, \nonumber \\
    &J_{\rm mf}^i = f_1 (\partial_i \mu + F_{0i}) +\dots, \nonumber \\
    &O_{\rm mf} = \tilde{\Gamma} \mu +\dots\,,
\end{align}
where the dots involve higher order terms in our perturbative scheme.

From \eqref{eq:frame}, we see that the combination $g_0+c_1$ is analogous to the new coefficient $\lambda$ identified in \cite{Armas:2021vku}, in the sense that it is an $m$-dependent proportionality constant between the chemical potential $\mu$ and the time derivative of the dynamical field $\phi$. Of course, in the case of spontaneous symmetry breaking, $\phi$ is a Goldstone mode and its relation to $\mu$ is a physical, dynamical equation of motion, \textit{i.e.} the Josephson relation. In the case of pure explicit breaking, all of the dynamics is captured by Eq. \eqref{eq:frame_cr}, while Eq. \eqref{eq:frame} is a trivial choice of hydrodynamic frame. Indeed, from \eqref{eq:app_l0}, we can easily see that only the combination of coefficients $\left(g_{0}+\Gamma_1+2c_1\right)$ enters in the dispersion relation. Despite $g_0+c_1$ appearing in a similar fashion as $\lambda$ of \cite{Armas:2021vku}, it pops up at a different order in terms of the explicit breaking scale $m$, and would appear as subleading in their analysis because of the different symmetry breaking pattern. We also note that, while $\lambda$ in \cite{Armas:2021vku} is dissipative,\footnote{This happens because $\lambda$ involves a combination of the new, non-dissipative coefficient denoted by $\sigma_\times$ in \cite{Armas:2021vku}, and the usual dissipative Goldstone diffusivity $\sigma_\phi$, which is absent in the purely explicit case.} here all the relevant coefficients $g_{0},\Gamma_1,c_1$ are non-dissipative.

In summary, in our purely explicit case, all these coefficients are at least $\mathcal{O}(m^2)$, and not order $\mathcal{O}(m)$ as those considered in \cite{Armas:2021vku}.\footnote{This is consistent with results for the linear axion model by Blaise Gout\'{e}raux and Ashish Shukla, to appear soon.} In presence of only explicit breaking, the possibility of having such a coefficients $\mathcal{O}(m)$ cannot be realized. On the contrary, whenever the additional spontaneous scale (the condensate) is introduced in the EFT description, then a combination of the spontaneous and explicit scales could bring such a term down to order $\mathcal{O}(m)$. In general, it is not surprising that the existence of additional scales allow for a richer set of possibilities in the scaling of the various coefficients. A similar situation happens for example in the case of translations for the pinning frequency and the Gell-Mann-Oakes-Renner relation, see \cite{Andrade:2017cnc,Andrade:2018gqk,Andrade:2020hpu}.

\subsection{Charge relaxation mode}\label{seczero}

In this section, we focus on the dispersion relation of the charge relaxation mode. In absence of any explicit breaking terms, such a mode corresponds to the diffusion of conserved charge and obeys the following dispersion relation
\begin{equation}
    \omega=-i \frac{\sigma}{\chi_{\rho\rho}}\,k^2+\dots
\end{equation}
where $\sigma$ is the electric conductivity and $\chi_{\rho\rho}\equiv \partial \rho/\partial \mu$ the charge susceptibility. In presence of soft explicit breaking of the $U(1)$ symmetry, this mode acquires a finite relaxation time at zero wave-vector and its dispersion relation becomes
\begin{equation}\label{aa}
    \omega=-i \Gamma-i D_q k^2+\cdots
\end{equation}
where $D_q$ does not anymore correspond to the standard charge diffusion $\sigma/\chi_{\rho\rho}$, as eventually will get corrected by terms proportional to the explicit breaking scale.
The charge relaxation mode is the lowest non-hydrodynamic mode in the quasi-normal modes (QNMs) spectrum, and it is expected to be well separated from the other microscopic degrees of freedom as long as the explicit breaking is kept small. Because of this reason, the relaxational late-time dynamics in that limit will be dominated by its dynamics.

With the EFT Lagrangian \eqref{L1} and \eqref{L2}, we can show \eqref{aa} and predict the relaxation rate $\Gamma$ and diffusion constant $D_q$. To this end, we turn off external sources and obtain the dispersion relation from $\delta S_{eff}/\delta \phi_a =0$ (setting $\phi_a=0$ in the equation)
\begin{align}
    0= i\omega a_1 + \omega^2 a_2 + k^2 a_3 +a_4 + a_5 i\omega k^2 + a_6 i \omega^3,  \label{dispersion_eq}
\end{align}
where for safety we have kept all third order derivatives. The various coefficients in the dispersion equation are expressed in terms of those in EFT action
\begin{align}
    &a_1 \equiv g_0 - c_1 + d_1 - \Gamma_1 = a_1^{(0)} + m^2 a_1^{(2)}, \nonumber \\
    &a_2 \equiv g_1 - c_3 +d_3 - \Gamma_4 = a_2^{(0)} + m^2 a_2^{(2)}, \nonumber \\
    &a_3 \equiv f_1 + h_0 + h_1 - c_4 - c_5 + d_4 + d_5 - \Gamma_3 = a_3^{(0)} + m^2 a_3^{(2)}, \nonumber \\
    &a_4 \equiv \tilde \Gamma = 0 + m^2 a_4^{(2)}, \nonumber \\
    & a_5 \equiv - g_3 - h_2 - h_3 + \Gamma_6 + c_7 + c_8 - d_7 - d_8 = a_5^{(0)} + m^2 a_5^{(2)},  \nonumber \\
    & a_6 \equiv -g_2 + \Gamma_5 + c_6 - d_6= a_6^{(0)} + m^2 a_6^{(2)}, \label{a1-6_EFT}
\end{align}
where we have expanded each coefficients to order $\mathcal O(m^2)$. In the hydrodynamic limit, the dispersion equation is solved and yields the dispersion shown in Eq.~\eqref{aa} with
\begin{align}
    \Gamma = - m^2 \frac{a_4^{(2)}}{a_1^{(0)}}, \qquad D_q = - \frac{a_3^{(0)}}{a_1^{(0)}} -  m^2 \left[ \frac{a_3^{(2)}}{a_1^{(0)}} - \frac{a_1^{(2)} a_3^{(0)}+a_4^{(2)} a_5^{(0)}}{(a_1^{(0)})^2}   - \frac{2 a_2^{(0)} a_3^{(0)} a_4^{(2)}}{(a_1^{(0)})^3}\right]. \label{Gamma_Dq_EFT}
\end{align}
Using the explicit holographic results in Eq.\eqref{holo_SK_coeffs}, along with the KMS conditions \eqref{eq:KMS_conditions}, we find
\begin{align}
    a_1^{(0)} & = 2r_h^2, \qquad a_1^{(2)} = \frac{1}{2} r_h^2 \log 2, \qquad a_2^{(0)}=0,\nonumber\\
    a_3^{(0)} & = -r_h, \qquad a_3^{(2)} = 0, \qquad a_4^{(2)} = - r_h^3,
\end{align}
while, from Ref.~\cite{deBoer:2018qqm}, we can read off the coefficient of the $k^4$ term in the dispersion relation of normal diffusion, implying that
\begin{align}
    a_5^{(0)} = -\frac{1}{2} \log 2.
\end{align}
Using these expressions in Eq.~\eqref{Gamma_Dq_EFT}, we immediately obtain
\begin{align}
    \Gamma = \frac{1}{2} m^2 r_h, \qquad \qquad D_q = \frac{1}{2r_h} + \mathcal{O}(m^4),\label{relaxation_rate_diffusion_SK}
\end{align}
where, surprisingly, the $\mathcal{O}(m^2k^2)$ term in the dispersion relation vanishes. In principle, this might be simply an artifact of the holographic model used, or of the probe limit that we assumed. However, there is also the possibility that such a result is a generic feature of the quasi-hydrodynamics structure revealed by holographic calculations. In order to validate this prospect, one should check other concrete models or prove it directly from a more careful field theory analysis. We plan to investigate this point further in the near future.

We will confirm the validity of the results above with two alternative methods. First, we will compute the dispersion relation using standard holographic perturbative techniques, which do not involve the SK contour. Second, we will perform a numerical computations for the QNMs of the field theory dual to our holographic model. As we will see, all the three methods give compatible results.

\subsubsection{Analytic computation of the lowest quasi-normal modes} \label{pert_ana_method}

In this section, we will consider the bulk action presented in Eq.~\eqref{Stueckelberg_action} and investigate the dispersion relation of the lowest non-hydrodynamic QNM, the charge relaxation mode.  Here, as a major difference with the computations presented in Section \ref{sec2}, we shall follow a fully on-shell formalism. More precisely, we will solve all the components of bulk EOMs for the fluctuations. For this purpose, we find it more convenient to work in the Schwarzschild coordinate system where the background line element is defined as
\begin{align}
    ds^2 = \frac{dr^2}{r^2f(r)} - r^2f(r) dt^2 + r^2 \delta_{ij} dx^i dx^j.
\end{align}
We denote the bulk fields in Schwarzschild coordinates using an extra tilde, $\tilde C_M$ and $\tilde \theta$. For convenience, we also assume the radial gauge, $\tilde C_r =0$. We decompose the bulk fields in Fourier space
\begin{align}
    \tilde C_M(r, x^\mu) = \int \frac{d\omega d k}{(2\pi)^2} \tilde C_M(r, \omega,k) e^{-i \omega t + i k x}, \qquad \qquad \tilde \theta(r, x^\mu) = \int \frac{d\omega dk}{(2\pi)^2} \tilde \theta(r, \omega,k) e^{-i \omega t + i k x}.
\end{align}
Since all bulk fields vanish in the background, the equations for the fluctuations correspond to those for the bulk fields themselves.

The transverse components of the gauge field $\tilde C_y$ or $\tilde C_z$ satisfy a single ordinary differential equation (ODE) given by
\begin{align}
    0 = \partial_r\left[ r^3f(r) \partial_r \tilde C_\alpha \right] + \frac{\omega^2}{rf(r)} \tilde C_\alpha - k^2 r^{-1} \tilde C_\alpha - m^2 r \tilde C_\alpha, \qquad \alpha =y, z.
\end{align}
Since we are mainly interested in the charge relaxation mode, we will only consider the fluctuations in the longitudinal sector and skip those in the transverse sector. We do not expect any (quasi-)hydrodynamic modes in the transverse sector.

The longitudinal modes $\{ \tilde C_t, \tilde C_x, \tilde \theta \}$ satisfy a system of linear ODEs which reads
\begin{align}
    0 = & \partial_r(r^3 \partial_r \tilde C_t) - \frac{1}{rf(r)}(k^2 \tilde C_t + \omega k \tilde C_x) - \frac{m^2 r}{f(r)} \tilde C_t - \frac{m^2 r i\omega}{f(r)} \tilde \theta, \nonumber \\
    0 = & \partial_r \left[r^3f(r) \partial_r \tilde C_x \right] + \frac{1}{rf(r)} (\omega^2 \tilde C_x + \omega k \tilde C_t) - m^2 r \tilde C_x + m^2 i k r \tilde \theta, \nonumber \\
    0 = & \partial_r \left[r^5f(r) \partial_r \tilde \theta\right] + \frac{r}{f(r)} (\omega^2 \tilde \theta - i\omega \tilde C_t) - r(k^2 \tilde \theta + ik \tilde C_x), \nonumber \\
    0 = & -i\omega \partial_r \tilde C_t - ik f(r) \partial_r \tilde C_x + m^2 r^2 f(r) \partial_r \tilde \theta, \label{eom_longit}
\end{align}
where the first three equations are dynamical, while the last one is a constraint. The constraint equation, when computed at the boundary, is related to the UV Ward identity. Here, it is important to notice that imposing any two of the three dynamical equations in \eqref{eom_longit}, plus the constraint equation, will automatically yield the remaining dynamical equation. Because of this reason, one of the three dynamical equations can be discarded.

Here, we start by clarifying the generic structure of solutions to the linear system in \eqref{eom_longit}. Near the horizon, the ingoing solutions behave as
\begin{align}
    &\tilde C_t^{\rm ig}(r, \omega,k) = [f(r)]^{1- i\omega/(4r_h)} \left[ \tilde C_t^h + \tilde C_t^1 (r-r_h) + \tilde C_t^2 (r-r_h)^2+ \cdots \right], \nonumber \\
    &\tilde C_x^{\rm ig}(r, \omega, k)= [f(r)]^{-i\omega/(4r_h)} \left[ \tilde C_x^h + \tilde C_x^1 (r-r_h) + \tilde C_x^2 (r-r_h)^2 + \cdots \right], \nonumber \\
    &\tilde \theta^{\rm ig}(r, \omega, k) = [f(r)]^{-i\omega/(4r_h)} \left[ \tilde \theta^h + \tilde \theta^1 (r-r_h) + \tilde \theta^2 (r-r_h)^2 + \cdots \right], \label{ig_solutions}
\end{align}
where $\tilde C_x^h$ and $\tilde \theta^h$ are two independent horizon data. In other words, all the rest coefficients $\tilde C_t^h$, $\tilde C_t^1$, $\tilde C_t^2$, $\tilde C_x^1$, $\tilde C_x^2$, $\tilde \theta^1$, $\tilde \theta^2$, $\cdots$ are fully determined in terms of the horizon data $\tilde C_x^h$ and $\tilde \theta^h$. For instance
\begin{align}
    \tilde C_t^h &= \frac{i k \tilde C_x^h - m^2 r_h^2 \tilde\theta^h}{4r_h - i\omega}, \nonumber \\
    \tilde C_x^1 &= \frac{\tilde C_x^h \left\{ 2m^2 r_h^2(4r_h - i\omega) + i \omega (-2k^2  -16 r_h^2 + 16 i\omega r_h +3 \omega^2) \right\} - 8 ik m^2 r_h^3 \tilde\theta^h}{4r_h^2(8r_h^2 -6i\omega r_h -\omega^2)}, \nonumber \\
    \tilde \theta^1 &= \frac{8 ik r_h \tilde C_x^h + \tilde \theta^h \left\{ k^2 (8r_h - 2 i\omega) +i \omega \left[ -2m^2 r_h^2 +3 \omega (4i r_h + \omega) \right] \right\}}{4r_h^2(8r_h^2 -6i\omega r_h -\omega^2)}.
\end{align}
Thus, the number of linearly independent ingoing solutions is two. These have to be supplemented by the pure gauge solution (labelled as ``pg'')
\begin{align}
    \tilde C_t^{\rm pg}(r, \omega, k) = -i \omega \tilde \xi, \qquad \tilde C_x^{\rm pg}(r, \omega, k) = ik \tilde \xi, \qquad \tilde \theta^{\rm pg}(r, \omega, k) = \tilde \xi,  \label{pg_solutions}
\end{align}
which is characterized by the residual gauge transformation parameter $\tilde \xi$.

Therefore, after imposing ingoing boundary conditions near the horizon (\textit{i.e.}, ruling out outgoing solutions), a generic solution for the linear system \eqref{eom_longit} can be written as
\begin{align}
    \tilde C_t(r, \omega, k) &= b_1 \tilde C_t^{\rm ig,\, I}(r, \omega,k)  + b_2 \tilde C_t^{\rm ig,\, II}(r, \omega,k) + b_3  \tilde C_t^{\rm pg}(r, \omega,k), \nonumber \\
    \tilde C_x(r, \omega, k) &= b_1 \tilde C_x^{\rm ig,\, I}(r, \omega,k) + b_2 \tilde C_x^{\rm ig,\, II}(r, \omega,k) + b_3  \tilde C_x^{\rm pg}(r, \omega,k), \nonumber \\
    \tilde \theta(r, \omega, k) &= b_1 \tilde \theta^{\rm ig,\, I}(r, \omega,k) + b_2 \tilde \theta^{\rm ig,\, II}(r, \omega,k) + b_3  \tilde \theta^{\rm pg}(r, \omega,k),  \label{generic_sol_CtCxtheta}
\end{align}
where the ingoing solution $\{\tilde C_t^{\rm ig,\, I},~ \tilde C_x^{\rm ig,\, I}, \, \tilde \theta^{\rm ig,\, I}  \}$ may be uniquely fixed by taking horizon data $\tilde C_x^h \neq 0, \tilde\theta^h =0$ in \eqref{ig_solutions}, and the other ingoing solution $\{\tilde C_t^{\rm ig,\, II},~ \tilde C_x^{\rm ig,\, II}, \, \tilde \theta^{\rm ig,\, II} \}$ to taking horizon data $(\tilde C_x^h = 0, \tilde\theta^h \neq 0)$ in \eqref{ig_solutions}. The coefficients $b_{1,2,3}$ will be fixed by the AdS boundary conditions. At this point, the original task of determining the QNM spectrum for the bulk theory reduces to working out the two ingoing solutions \eqref{ig_solutions}. We will work perturbatively in the bulk mass $m$ so that
\begin{align}
    \tilde C_\mu = \tilde C_\mu^{(0)} + m^2 \tilde C_\mu^{(2)} + \cdots, \qquad \qquad \tilde \theta = \tilde \theta^{(0)} + \cdots.
\end{align}
Additionally, at each order in the $m$-expansion above, we will work in the hydrodynamic limit, say $\omega, k \ll T$. The QNM spectrum corresponds to solving the following expression\footnote{Here, we shall subtract the $\log r$-terms at the AdS boundary.}
\begin{equation}
\left |\begin{array}{cccc}
\tilde C_t^{\rm ig, I}(\infty) & \tilde C_t^{\rm ig, II}(\infty)   & \tilde C_t^{\rm pg}(\infty) \\
\tilde C_x^{\rm ig, I}(\infty) & \tilde C_x^{\rm ig, II}(\infty) & \tilde C_x^{\rm pg}(\infty)  \\
\tilde \theta^{\rm ig, I}(\infty) & \tilde \theta^{\rm ig, II}(\infty) & \tilde \theta^{\rm pg}(\infty) \\
\end{array}\right|=0. \label{dete}
\end{equation}
Recall that our goal is to obtain Eq.~\eqref{aa} using  holographic calculations. However, the presence of a gap ($\sim \Gamma$) makes the calculations subtle, which requires \eqref{dete} to properly cover third order derivatives in the dispersion equation \eqref{dispersion_eq}. After careful examination, we find this will be achieved if one covers second order derivative corrections in each ingoing solutions.

Without presenting more details, we summarize the main results for the two ingoing solutions \eqref{ig_solutions}. For the first ingoing solution, we take the horizon data as $(\tilde C_x^h, \tilde \theta^h) = (1,0)$ and find that
\begin{align}
    &\tilde C_t^{\rm ig,\, I}(r, \omega,k) = \left[f(r)\right]^{1- i\omega/(4r_h)} \left[ \frac{ik r^2}{2r_h(r^2 +r_h^2)} + \cdots + m^2 \tilde C_t^{\rm ig, I(2)(1)} + \cdots \right], \nonumber \\
    &\tilde C_x^{\rm ig, \, I}(r, \omega,k) = \left[f(r)\right]^{- i\omega/(4r_h)} \left[ 1 + \frac{i\omega}{2r_h} \log \frac{r^2 + r_h^2}{2r^2} + \cdots + m^2 \tilde C_x^{\rm ig, I(2)(1)} + \cdots \right], \nonumber \\
    &\tilde \theta^{\rm ig, \, I}(r, \omega, k) = \left[ f(r)\right]^{- i \omega/(4r_h)} \left[ - \frac{i k}{4r_h^2} \log \frac{r^2 +r_h^2}{2r^2} + \cdots + \mathcal{O}(m^2) \right], \label{1st_ingoing_sol}
\end{align}
where
\begin{align}
&\tilde C_t^{\rm ig, I(2)(1)} =\frac{ik r^2}{8r_h(r_h^4 - r^4)} \left[ r^2 -r_h^2 +(r^2 +r_h^2) \log\frac{2r_h^2}{r^2 +r_h^2} \right], \\
   & \tilde C_x^{\rm ig, I(2)(1)} =\frac{1}{4} \log \frac{r^2 + r_h^2}{2r_h^2} - \frac{i\omega}{4r_h} \log r + \frac{i \omega}{48r_h} (\pi^2 + 6\log^2 2 + 12\log 2 \log r_h) + \mathcal{O}(r^{-2}), \nonumber
\end{align}
and the second term is expressed only near the boundary, in the limit $r \to \infty$. The omitted terms denoted by $\cdots$ in \eqref{1st_ingoing_sol} represent second order derivatives, which are too lengthy to be presented. However, some of them are crucial in correctly producing \eqref{dispersion_eq}.

Near the AdS boundary, the first ingoing solution behaves as
\begin{align}
    \tilde C_t^{\rm ig, \, I}(r \to \infty) = & m^2 \frac{ik}{4r_h} \log r + \frac{ik}{2r_h} + \frac{\log 2}{4r_h^2} \omega k - m^2 \frac{ik}{8r_h} \left[ 1+ \log(2r_h^2) \right] \nonumber \\
    & - m^2 \frac{\pi^2 + 12 \log^22}{384r_h^2} \omega k + \mathcal{O}(r^{-1}), \nonumber \\
    \tilde C_x^{\rm ig, \, I}(r \to \infty) = & m^2\left( \frac{1}{2} - \frac{i\omega \log2}{4r_h} \right) \log r + 1 - \frac{i\omega}{2r_h} \log 2 - \frac{\pi^2 + 6 \log^22}{48r_h^2} \omega^2 \nonumber \\
    &  - m^2 \frac{1}{4} \log(2r_h^2)  + m^2 \frac{i\omega}{48r_h} \left( \pi^2 + 6\log^22 +12 \log 2 \log r_h \right) \nonumber \\
    & + \#_1 \omega^2 + \#_2 q^2 + \mathcal{O}(r^{-1}), \nonumber \\
    \tilde \theta^{\rm ig, \, I}(r \to \infty) = & \frac{ik}{4r_h^2} \log2 - \frac{\pi^2 - 24(\log2 -1) \log2}{192r_h^2} \omega k + \mathcal{O}(r^{-1}), \label{1st_ingoing_bdy}
\end{align}
where $\#_1, \#_2$ will play no role in subsequent calculation of QNM.

For the second ingoing solution, we take the horizon data as $(\tilde C_x^h, \tilde \theta^h) = (0,1)$ and find that
\begin{align}
    &\tilde C_t^{\rm ig,\, II}(r, \omega,k) = \left[f(r)\right]^{1- i\omega/(4r_h)} m^2 \left[- \frac{r_h r^2}{2(r^2+ r_h^2)} + \tilde C_t^{\rm ig, II(2)(1)} + \cdots \right], \nonumber \\
    &\tilde C_x^{\rm ig, \, II}(r, \omega,k) = \left[f(r)\right]^{- i\omega/(4r_h)} m^2 \left[ - \frac{1}{4} i k \log \frac{r^2 +r_h^2}{2r_h^2} + \cdots \right], \nonumber \\
    &\tilde \theta^{\rm ig, \, II} = \left[f(r)\right]^{- i\omega/(4r_h)} \left[ 1 + \cdots + \mathcal{O}(m^2) \right], \label{2nd_ingoing_sol}
\end{align}
where
\begin{equation}
    \tilde C_t^{\rm ig, II(2)(1)}=\frac{ i \omega r^2}{4(r^4- r_h^4)} \left[ (r^2 -r_h^2) \left(\log \frac{r^2}{r_h^2} - 1\right) + r_h^2 \log\frac{r^2 + r_h^2}{2r_h^2} \right].
\end{equation}
Here, as in \eqref{1st_ingoing_sol}, the terms denoted by $\cdots$ in \eqref{2nd_ingoing_sol} represent second order derivatives and are too lengthy to be presented above. Similarly, they are important for correctly covering third order terms in the dispersion equation \eqref{dispersion_eq}.

Near the AdS boundary, the second ingoing solution behaves as
\begin{align}
    \tilde C_t^{\rm ig, \, II}(r \to \infty) = & m^2 \frac{1}{2} i\omega \log r - m^2 \frac{r_h}{2} - m^2 \frac{1}{4} i\omega (1+ 2 \log r_h) - m^2 \frac{\pi^2}{192 r_h} \omega^2 \nonumber \\
    & - m^2 \frac{\log 2}{8r_h} q^2 + \mathcal{O}(r^{-1}), \nonumber \\
    \tilde C_x^{\rm ig, \, II}(r \to \infty) = & - m^2 \frac{1}{2} ik \log r + m^2 \frac{1}{4} ik \log(2r_h^2) + m^2 \frac{\pi^2 + 4\log^2 2}{64r_h} \omega k + \mathcal{O}(r^{-1}), \nonumber \\
    \tilde \theta^{\rm ig, \, II}(r \to \infty) = & 1 + \frac{-\pi^2 + 12(\log2 -2) \log2}{96r_h^2} \omega^2 + \frac{\log 2}{4r_h^2} q^2 + O(r^{-1}). \label{2nd_ingoing_bdy}
\end{align}

With \eqref{1st_ingoing_bdy} and \eqref{2nd_ingoing_bdy}, it is straightforward to solve \eqref{dete} in the hydrodynamic limit. As expected, we  obtain the dispersion relation \eqref{aa} with
\begin{align}
    \Gamma = \frac{r_h}{2} m^2, \qquad D_q = \frac{1}{2r_h} +  \mathcal{O} (m^4). \label{Gamma_Dq_final}
\end{align}
As anticipated in section 
\ref{seczero}, we find the diffusion constant does not receive correction at the order $\mathcal O(m^2)$. As already mentioned, further studies are needed to ascertain whether this property is a reflection of a deeper, and maybe even universal, underlying structure. This result will be further confirmed through numerical calculation in next section.
\begin{figure}[htp]
    \centering
    \includegraphics[width=0.475 \linewidth]{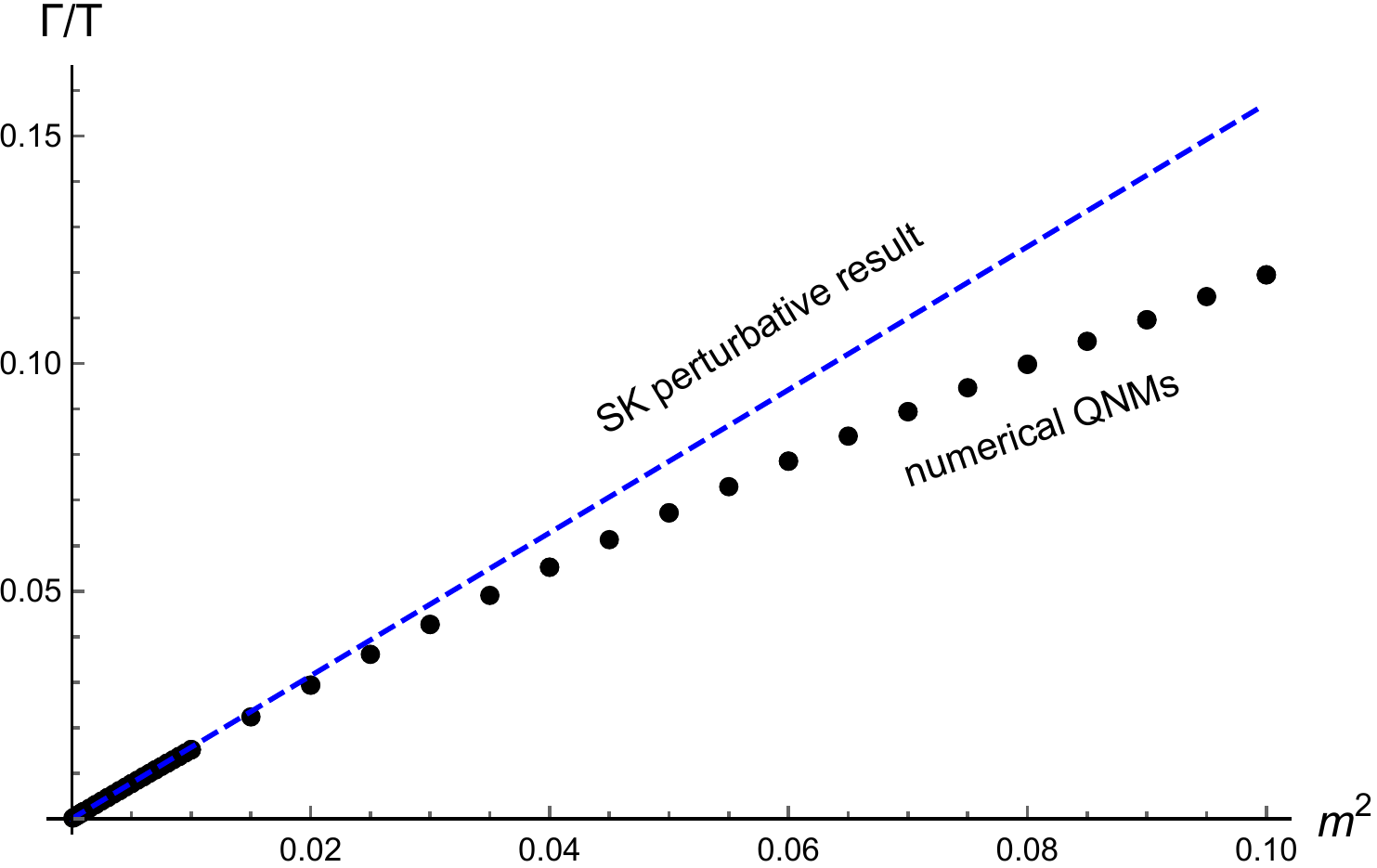}\quad
    \includegraphics[width=0.475 \linewidth]{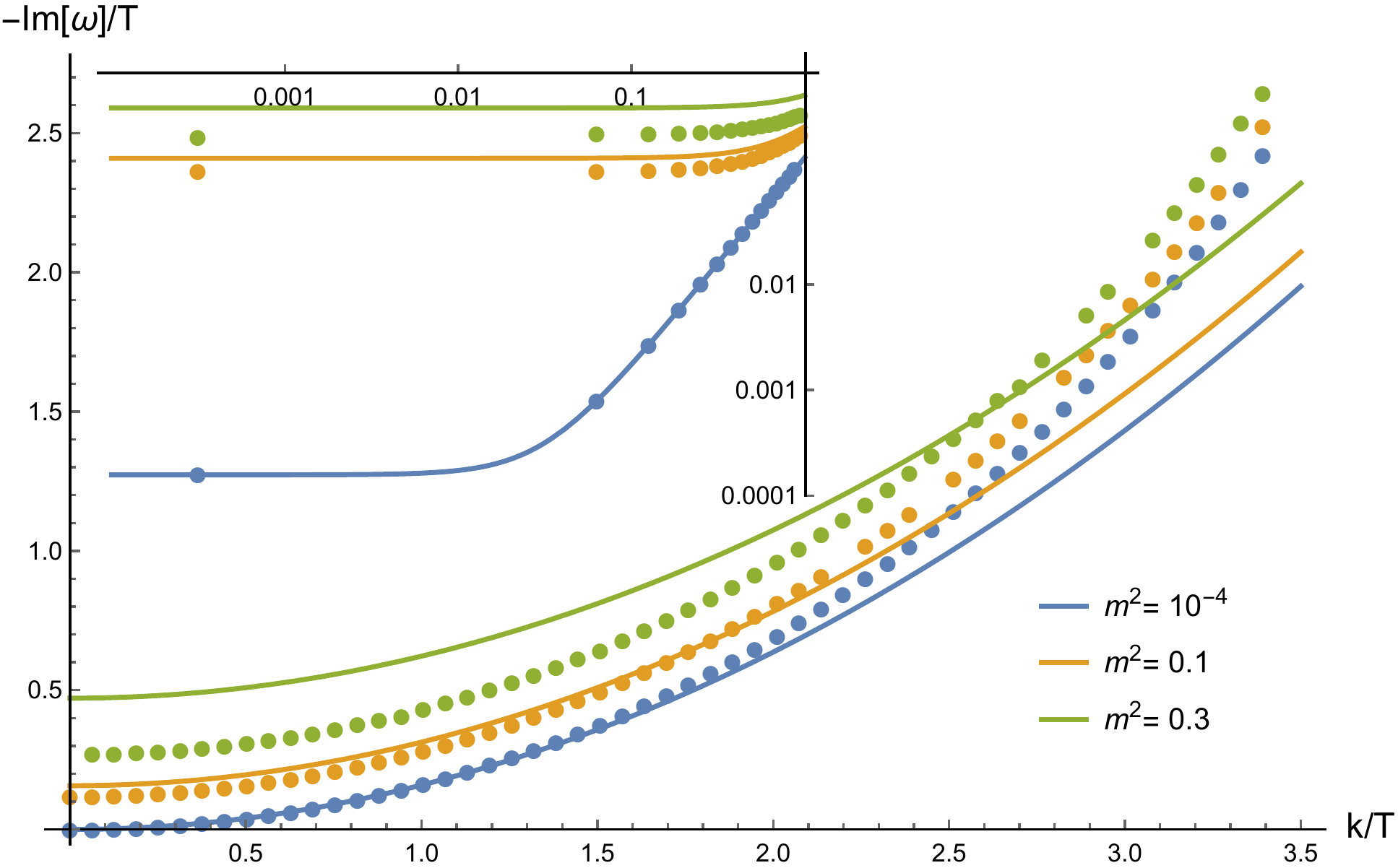}
    \caption{\textbf{Left:} The charge relaxation rate extracted from the imaginary part of the lowest QNM at $k=0$ (black symbols) vs. the prediction from the perturbative SK computation in Eq.~\eqref{Gamma_Dq_final}. \textbf{Right:} The dispersion relation of the lowest non-hydrodynamic mode for different values of the dimensionless Proca mass. The solid lines are the SK perturbative results in Eq.~\eqref{Gamma_Dq_final}. The inset shows a zoom for small values of the wave-vector $k$.}
    \label{fig3}
\end{figure}
\subsubsection{Numerical computation} \label{numeric_study}
The analytical computation presented in section \ref{pert_ana_method} is perturbative in the bulk mass $m$, the frequency $\omega$ and the wave-vector $k$. In other words, the result in \eqref{aa} and \eqref{Gamma_Dq_final} reproduces only the dispersion relation of the lowest QNM in the hydrodynamic limit, and in the limit of small explicit breaking, $m/T \ll 1$. The complete QNMs spectrum beyond such approximations can nevertheless be obtained using numerical methods. The idea is exactly the same as that in the previous section, but the EOMs \eqref{eom_longit} will be now solved numerically using the same boundary conditions. In this way, we can search for the solutions of \eqref{dete} at any order in $\omega,k,m$. For simplicity, we focus only on the lowest QNM, \textit{i.e.}, the charge relaxation mode. First, we extract its imaginary part at zero wave-vector as a function of the explicit breaking parameter $m$. The result is shown in the left panel of Figure \ref{fig3}. As expected, for small values of the explicit breaking parameter $m/T$, the perturbative result in Eq.~\eqref{Gamma_Dq_final} is in good agreement with the numerical data. Beyond a critical value, the numerical results start to deviate from the perturbative formula (first equation of \eqref{Gamma_Dq_final}), suggesting that the neglected higher order corrections become important. In the right panel of Figure \ref{fig3}, we show the complete dispersion relation of the charge relaxation mode for different values of $m/T$. For small values of that parameter we see that Eq.~\eqref{Gamma_Dq_final} is in good agreement with the data up to $k/T \approx 1$, within the hydrodynamic limit. By increasing the value of the explicit breaking parameter, we observe that the perturbative result deviates from the numerical data. In summary, the numerical computation confirms the validity of our results both in the SK formalism and in the single AdS perturbative scheme.

One of the main results of the previous analysis is that the quadratic term in the dispersion will be not affected at order $m^2$, but only at higher order. In order to check this with the numerics, we focus on the low mass regime, $m/T \ll 1$, and carefully track the quadratic term in the dispersion relation.

\begin{figure}[h]
    \centering
    \includegraphics[width=0.6\linewidth]{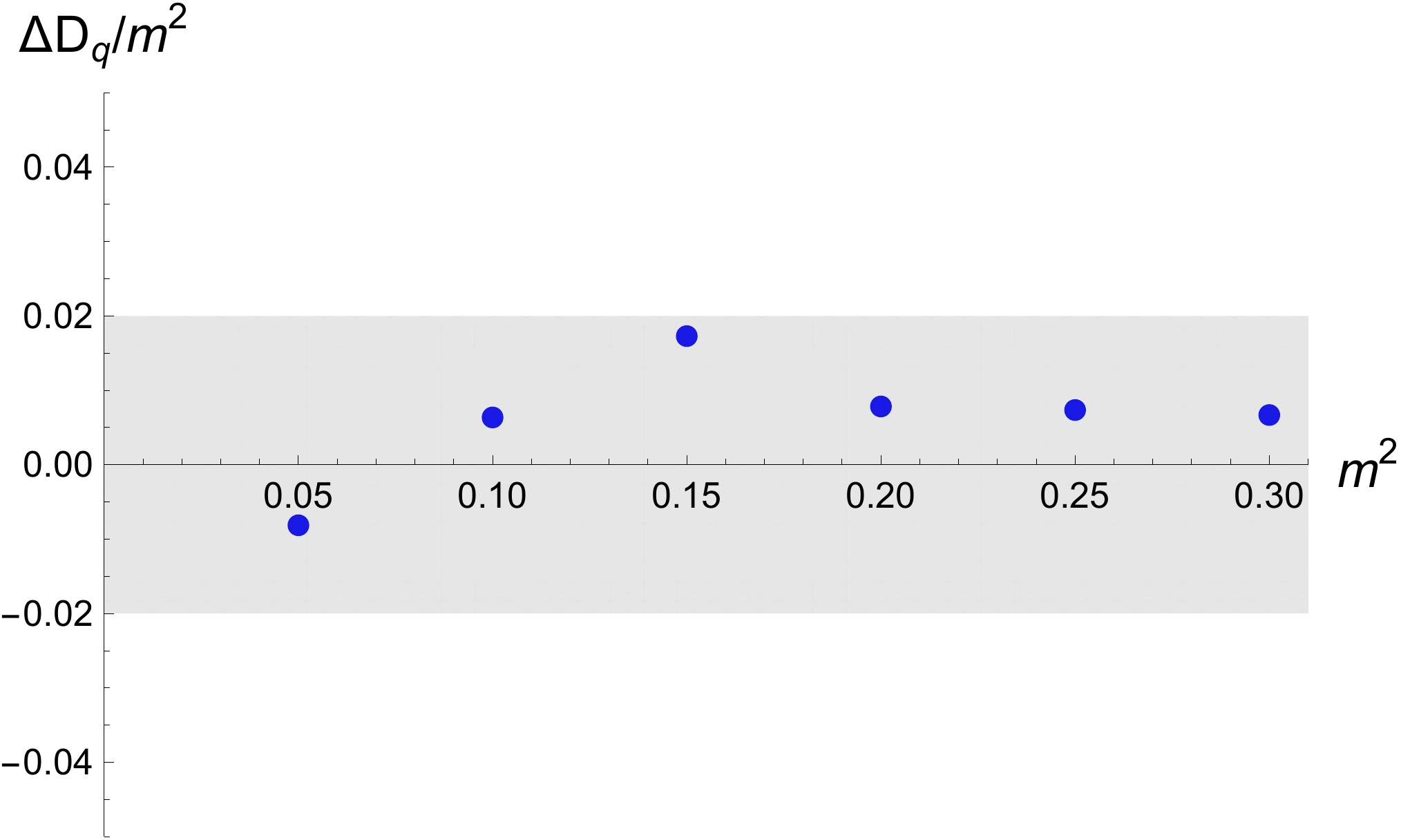}
    \caption{The deviation of the quadratic coefficient in the dispersion relation from its zero mass value, $\Delta D_q$ as a function of the mass squared.}
    \label{fig:proof}
\end{figure}

In Figure \ref{fig:proof}, we consider the quadratic term $\sim k^2$ in the dispersion relation of the relaxing charge fluctuations. We subtract from it the value at zero mass, and carefully look at the corrections at order $m^2$. We define the variation of such a coefficient from the zero mass value $\Delta D_q$ and plot it, after normalizing it by $m^2$, as a function of the mass itself. For small values of $m^2$, we see that the resulting numerical data are scattered in a region $\pm 2 \%$ around zero. Therefore, within that confidence, we can conclude that the $\mathcal{O}(m^2)$ correction to the quadratic coefficient is compatible with zero, as derived from the previous theoretical analysis. Our numerics show that, if such a correction existed, it would be extremely small, of order $10^{-2}$, which is rather unnatural from the EFT perspective. 

\section{Summary and Outlook} \label{sec4}

In this work, we have initiated the study of quasi-hydrodynamics using holographic Scwhinger-Keldysh methods. We have focused on the simplest example of a system with a global $U(1)$ symmetry which is explicitly broken in a controllable way, and whose breaking can be made parametrically small. In order to do that, we have considered a holographic massive vector Proca model, where the bulk mass $m$ is responsible for the breaking of the $U(1)$ symmetry. We take this as the simplest example of relaxed hydrodynamics. Using holographic SK techniques, we have derived the low-energy effective action of the boundary field theory in presence of dissipation and finite temperature effects, in the limit of small breaking $m/T \ll 1$, \textit{i.e.}, the quasi-hydrodynamic regime. The results obtained within holographic SK approach are compared and confirmed using standard EFT methods, already presented in \cite{Delacretaz:2021qqu,Armas:2021vku}, and the analytical and numerical computations of the lowest QNMs. We find perfect agreement between all these different methods. Interestingly, we find that the term $\mathcal{O}(m^2k^2)$ in the dispersion of the damped charge diffusion mode vanishes identically. It would thus be interesting to determine whether that is just an accident of the model considered or a manifestation of a more fundamental, and perhaps universal, characteristic of the quasi-hydrodynamics structure.

Among the various results, we have formally derived the field theory Ward identity for the $U(1)$ current from the holographic Proca model. We have confirmed that its form is in agreement with the EFT predictions, and with the form previously assumed in the literature, \textit{e.g.}, \cite{Ammon:2021pyz,Jimenez-Alba:2015awa} (see \cite{Klebanov:2002gr} for a previous formal derivation in the context of string theory). Taking advantage of the derivation of the boundary effective SK action, we have directly observed the appearance of additional transport coefficients, which are analogous to those proposed in \cite{Armas:2021vku}, appearing in the case of pseudo-spontaneous symmetry breaking. By direct analytical computation, we have shown that terms of that kind naturally appear in our effective action. Nevertheless, in the case of explicit breaking, they always appear at quadratic order in the explicit breaking scale, indicated as $m$ in our analysis, and not at linear order, as for the terms described in \cite{Armas:2021vku}. This difference in the orders is not a coincidence, nor a result of the different counting scheme chosen, but rather a direct manifestation of the different symmetry breaking pattern; in presence of only explicit breaking, it is impossible to realize those coefficients at linear order.

From a more practical point of view, even though we are not aware of any realistic system where a vector $U(1)$ symmetry is explicitly broken, our setup could serve as a toy model for several physical scenarios which are of great interest in different fields of research, and they are all connected by the same quasi-hydrodynamic structure. A first important case is that of axial charge relaxation in the context of QCD and condensed matter systems with anomalous transport. In QCD, for example, axial symmetry is broken by gluonic effects, which can be parameterized in holography using an explicit mass for the bulk gauge field \cite{Klebanov:2002gr} (see for example \cite{Jimenez-Alba:2015awa,Jimenez-Alba:2014iia}). A similar situation appears in the context of relativistic spin hydrodynamics, where the spin current is not conserved, but at least for massive fermions, its relaxation can be made parametrically long by taking the non-relativistic limit of large fermion mass \cite{Hongo:2022izs}. This is as well the case for magneto-hydrodynamics, in which the global $U(1)$ electric symmetry is broken, and where both electric field and charge fluctuations relax in time \cite{Grozdanov:2016tdf,Hernandez:2017mch} (see \cite{Grozdanov:2017kyl,Poovuttikul:2021fdi,Ahn:2022azl} for the holographic realizations).

Additionally, any system close to a critical point also displays a similar behavior, since the fluctuations of the amplitude of the order parameter relax very slowly close to criticality. This has been formalized in the context of relativistic hydrodynamics with the so-called Hydro+ formalism \cite{Stephanov:2017ghc}, but it appears more general in the study of critical phenomena \cite{RevModPhys.49.435} (\textit{e.g.}, the amplitude mode in superfluids/superconductors \cite{Donos:2022xfd,Donos:2023ibv}). Moreover, several hydrodynamic models (\textit{e.g.}, Maxwell's model \cite{maxwell1867iv}, Cattaneo's model \cite{Cattaneo1958431}, Israel-Stewart model \cite{ISRAEL1979341}, and generalizations \cite{Gavassino:2022roi}) involve at least one non-hydrodynamic mode (which is necessary in relativistic theories to preserve causality), and therefore fall in the more general class of quasi-hydrodynamics. In the context of classical liquids, the quasi-hydrodynamic structure is not a theoretical framework but it is responsible for several physical phenomena, such as the onset of gapped shear waves, which can be observed experimentally \cite{nosenko2006}. Finally, a widely studied framework where quasi-hydrodynamics is at work is that of momentum relaxation in condensed matter systems, where disorder, impurities, or simply Umklapp scattering produce a dissipation of momentum (\textit{e.g.}, the Drude model). 

We believe that extending our computation to these more complex cases would be very fruitful in the near future, not only to understand better the validity and the meaning of quasi-hydrodynamics but also to reveal in depth the structure of the low-energy effective descriptions for such systems. We are confident that holography might always be a good partner in these explorations, and we are planning to consider some of these problems in the near future.

\acknowledgments{We would like to thank Navid Abbasi, Danny Brattan, Aristomenis Donos, Paolo Glorioso, Blaise Gout\'{e}raux,  Sebastian Grieninger, Michael Landry and Ashish Shukla for useful discussions. We would also like to thank Akash Jain and Jay Armas for discussions on the relation between our results and theirs. M.B. acknowledges the support of the Shanghai Municipal Science and Technology Major Project (Grant No.2019SHZDZX01) and the sponsorship from the Yangyang Development Fund. V.Z. is supported by the European Research Council (ERC) under the European Union’s Horizon 2020 research and innovation program (grant agreement No758759).}

\appendix

\section{Details of the holographic Schwinger-Keldysh computations}\label{app1}

In this appendix, we present the perturbative solutions for bulk fields. When the source terms are ignored, all the EOMs become homogeneous ODEs, which can be solved analytically
\begin{align}
    &C_v^h(r) = c_v^1 + \frac{c_v^2}{r^2},\quad C_i^h(r) = c_i^1 + c_i^2 \log \frac{r^2 - r_h^2}{r^2 + r_h^2},\quad  \theta^h(r) = c_\theta^1 + c_\theta^2 \log \frac{r^4 - r_h^4}{r^4}. \label{linear_indep_solutions}
\end{align}
Obviously, the two linearly independent solutions for $C_v$ are regular over the entire contour. This is one of the reason why the extra condition \eqref{Cv=0} has to be imposed at the horizon. Thus, $C_v$ will have a piecewise solution at each order. Later on, the linearly independent solutions for $C_i$ and $\theta$ will be used to build Green's functions on the radial contour of Figure \ref{holographic_SK_contour}.

\subsection*{The solutions for $C_\mu^{(0)(1)}$ and $\theta^{(0)(0)}$}

The solution for $C_v^{(0)(1)}$ is piecewise
\begin{align}
    &C_v^{(0)(1)}(r) = B_{1v} \left(1- \frac{r_h^2}{r^2} \right), \qquad r \in [r_h - \epsilon, \infty_1), \nonumber \\
    &C_v^{(0)(1)}(r) = B_{2v} \left(1- \frac{r_h^2}{r^2} \right), \qquad r \in [r_h - \epsilon, \infty_2). \label{Cv01}
\end{align}
The solution for $C_i^{(0)(1)}$ is
\begin{align}
    C_i^{(0)(1)}(r) = B_{2i} + \frac{B_{ai}}{2i\pi} \log \frac{r^2-r_h^2}{r^2 + r_h^2}.  \label{Ci01}
\end{align}
The solution for $\theta^{(0)(0)}$ is
\begin{align}
    \theta^{(0)(0)}(r) = \vartheta_2 + \frac{\vartheta_a}{2i\pi} \log\frac{r^4-r_h^4}{r^4} . \label{theta00}
\end{align}

It is straightforward to check that all the lowest order solutions satisfy the respective EOMs with correct boundary conditions. It is important to stress that, in contrast to $C_v^{(0)(1)}$ which is regular over the entire contour, both $C_i^{(0)(1)}$ and $\theta^{(0)(0)}$ contain a singular part, which roughly behaves as $\log(r-r_h)$ near the horizon, which is a multi-valued function.

Near the AdS boundaries, we read off the normalizable modes for the lowest order solutions
\begin{align}
    &\tilde J_{1v}^{(0)(1)} = -r_h^2 B_{1v},  \qquad \qquad \qquad  \tilde J_{2v}^{(0)(1)} = -r_h^2 B_{2v} ,\nonumber \\
    &\tilde J_{1i}^{(0)(1)} = \frac{i}{\pi} r_h^2 B_{ai},  \qquad \qquad \qquad  \tilde J_{2i}^{(0)(1)} = \frac{i}{\pi} r_h^2 B_{ai}, \nonumber \\
    &\tilde O_1^{(0)(0)} = \frac{i r_h^4}{2\pi} \vartheta_a, \qquad \qquad \qquad \quad \tilde O_2^{(0)(0)} = \frac{i r_h^4}{2\pi} \vartheta_a,
\end{align}
which, via \eqref{Seff_formal}, give the leading-order result for the effective action. Here, we record the explicit results for each of the terms in \eqref{Seff_formal}
\begin{align}
     &B_\mu \tilde J^{\mu (0)(1)}\bigg|_2^1 = 2r_h^2 B_{av} B_{rv} + \frac{ir_h^2}{\pi} B_{ai} B_{ai}, \nonumber \\
     &\frac{1}{2} m^2 B_\mu \tilde J^{\mu (0)(1)}\bigg|_2^1 = \frac{1}{2} m^2 \left( 2r_h^2 B_{av} B_{rv} + \frac{ir_h^2}{\pi} B_{ai} B_{ai} \right), \nonumber \\
     &2m^2 \vartheta \tilde O^{(0)(0)} \bigg|_2^1 = m^2 \frac{i r_h^4}{\pi} \vartheta_a \vartheta_a, \nonumber \\
     &-\frac{3}{8}m^2 \tilde J^{\mu (0)(1)} \partial_\mu \vartheta \bigg|_2^1 = - \frac{3}{8}m^2 \left[ r_h^2 B_{av} \partial_v \vartheta_r + r_h^2 B_{rv} \partial_v \vartheta_a + \frac{i}{\pi} r_h^2 B_{ai} \partial_i \vartheta_a \right],
\end{align}
where the first and third lines are in perfect agreement with those of \cite{Glorioso:2018mmw}. Here, the diffusive field $\phi$ appears explicitly, representing the effect of explicit symmetry breaking. The last line represents a renormalization scheme-dependent contribution.

\subsection*{Solutions for $C_v^{(0)(2)}$ and $C_v^{(2)(1)}$ }

Beyond the lowest order, the EOM for $C_v^{(n)(l)}$ can be solved via direct integration. To this end, we examine the behavior of the sources near the AdS boundary and horizon
\begin{align}
    &j_v^{(0)(2)} (r \to \infty_s) = - \partial_v B_{sv} + \mathcal{O}(r^{-2}), \nonumber \\
    &j_v^{(0)(2)} (r \to r_h) = - \frac{r_h } {r-r_h} \partial_v B_{sv} + \cdots, \nonumber \\
    &j_v^{(2)(1)}(r \to \infty_s) = r (B_{sv} - \partial_v \vartheta_{s} )- \frac{r_h^2}{r} B_{sv} + \mathcal{O}(r^{-3}), \nonumber \\
    & j_v^{(2)(1)} (r \to r_h) = \#_1\frac{\log (r-r_h)}{r-r_h} + \frac{\#_1}{r-r_h},
\end{align}
where the leading order terms near the AdS boundary essentially correspond to those in the near-boundary expansion of $C_v^{(0)}$ and $C_v^{(1)}$. The divergent behavior near the horizon prohibits integrating the EOMs from the horizon. Similarly, the near-boundary behavior implies that we cannot just perform the integration from the AdS boundaries. To circumvent this problem, we redefine the bulk fields as
\begin{align}
    &C_v^{(0)(2)}(r) = \tilde C_v^{(0)(2)}(r) + \frac{\partial_v B_{sv}}{r}, \nonumber \\  &C_v^{(2)(1)}(r) = \tilde C_v^{(2)(1)}(r) + \frac{1}{2} (B_{sv} - \partial_v \vartheta_s) \log r + \frac{r_h^2 \log r}{2r^2} B_{sv},
\end{align}
so that the relevant EOMs are modified into
\begin{align}
    \partial_r \left[ r^3 \partial_r \tilde C_v^{(0)(2)}(r) \right] = \tilde j_v^{(0)(2)}(r), \qquad \qquad  \partial_r \left[ r^3 \partial_r \tilde C_v^{(2)(1)}(r) \right] = \tilde j_v^{(2)(1)}(r),
\end{align}
where the new sources are
\begin{align}
    &\tilde j_v^{(0)(2)}(r) = j_v^{(0)(2)}(r) + \partial_v B_{sv}, \nonumber \\
    &\tilde j_v^{(2)(1)}(r) = j_v^{(2)(1)}(r) -r (B_{sv} - \partial_v \vartheta_s) + \frac{r_h^2}{r} B_{sv},
\end{align}
which now behave well near the AdS boundaries. Then, the solutions are
\begin{align}
    &\tilde C_v^{(0)(2)}(r) = \int_{\infty_s}^r \left[ \frac{1}{x^3} \int_{\infty_s}^x \tilde j_v^{(0)(2)}(y) dy + \frac{c_s^{(0)(2)}}{x^3} \right] dx, \nonumber \\
    &\tilde C_v^{(2)(1)}(r) = \int_{\infty_s}^r \left[ \frac{1}{x^3} \int_{\infty_s}^x \tilde j_v^{(2)(1)}(y) dy + \frac{c_s^{(2)(1)}}{x^3} \right] dx,
\end{align}
which should be understood as piecewise, and the lower bound $\infty_s$ helps to distinguish between solution on the lower branch and that on the upper branch. The remaining integration constants are fixed by the vanishing horizon conditions for $C_v$
\begin{align}
    &c_s^{(0)(2)} = 2r_h^2 \int_{\infty_s}^{r_h} \left[ \frac{1}{x^3} \int_{\infty_s}^x \tilde j_v^{(0)(2)}(y) dy \right] dx + 2r_h \partial_v B_{sv}, \nonumber \\
    &c_s^{(2)(1)} = 2r_h^2 \int_{\infty_s}^{r_h} \left[ \frac{1}{x^3} \int_{\infty_s}^x \tilde j_v^{(2)(1)}(y) dy \right] dx + (2 B_{sv} - \partial_v \vartheta_s) r_h^2 \log r_h.
\end{align}
Then, the normalizable modes are determined in terms of integration constants
\begin{align}
    \tilde J_{sv}^{(0)(2)} = - \frac{1}{2} c_s^{(0)(2)}, \qquad \qquad  \tilde J_{sv}^{(2)(1)} = - \frac{1}{2} c_s^{(2)(1)}, \qquad \qquad s=1~{\rm or}~2.
\end{align}
For later convenience, we record the explicit expressions for $C_{sv}^{(0)(2)}$ and $C_{sv}^{(2)(1)}$ (with $s=1,2$)
\begin{align}
    &C_{sv}^{(0)(2)}(r) = \frac{\partial_v B_{sv}}{4r_h} \left( 1- \frac{r_h^2}{r^2} \right) \left[ \pi - 2 \arctan\left( \frac{r}{r_h}\right) + \log \frac{r+r_h}{r-r_h} \right], \nonumber \\
    &C_{sv}^{(2)(1)}(r) = - \frac{r_h^2 \log (2r_h^2)}{2r^2} B_{sv} + \frac{1}{4}B_{sv} \left( 1+ \frac{r_h^2}{r^2}\right) \log (r^2+ r_h^2) + \cdots,  \label{Cv02_Cv21_sol}
\end{align}
where the $\cdots$ denotes $\vartheta$-terms that are too lengthy to be written here. The normalizable modes of $C_{sv}^{(0)(2)}$ and $C_{sv}^{(2)(1)}$ are
\begin{align}
    \tilde J_{1v}^{(0)(2)} = & \tilde J_{2v}^{(0)(2)} =0, \nonumber \\
    \tilde J_{1v}^{(2)(1)} = & \left[ \frac{i r_h^2}{8\pi} (2\log 2 - \log^2 2) + \frac{r_h^2}{8} \log(2r_h^2) \right] \partial_v \vartheta_a  + \frac{r_h^2}{4} \log(2r_h^2)\, \partial_v \vartheta_r \nonumber \\
    & - \frac{r_h^2}{8} \left[ -1 + 2\log(2r_h^2) \right] B_{av} - \frac{r_h^2}{4} \left[ -1 + 2\log(2r_h^2) \right] B_{rv}, \nonumber \\
    \tilde J_{2v}^{(2)(1)} = & \left[ \frac{i r_h^2}{8\pi} (2\log 2 - \log^2 2) - \frac{r_h^2}{8} \log(2r_h^2) \right] \partial_v \vartheta_a  + \frac{r_h^2}{4} \log(2r_h^2)\, \partial_v \vartheta_r \nonumber \\
    & + \frac{r_h^2}{8} \left[ -1 + 2\log(2r_h^2) \right] B_{av} - \frac{r_h^2}{4} \left[ -1 + 2\log(2r_h^2) \right] B_{rv},
\end{align}
which yield
\begin{align}
   & B_v \tilde J^{v(0)(2)}\bigg|_2^1 = 0, \nonumber \\
   & \frac{1}{2} m^2 B_v \tilde J^{v(0)(2)}\bigg|_2^1 = 0, \nonumber \\
   & m^2 B_v \tilde J^{v(2)(1)} \bigg|_2^1 = m^2 \frac{ir_h^2}{8\pi} (\log^2 2 - 2 \log 2) B_{av} \partial_v \vartheta_a + m^2 \frac{r_h^2}{2} \left[ 2 \log (2r_h^2) -1 \right] B_{av} B_{rv}  \nonumber \\
   & \qquad \qquad \qquad \qquad - m^2 \frac{r_h^2}{4} \log (2r_h^2) \left( B_{av} \partial_v \vartheta_r + B_{rv} \partial_v \vartheta_a \right), \nonumber \\
   & \frac{3}{8} m^2 \tilde J^{v(0)(2)} \partial_v \vartheta \bigg|_2^1 =0.
\end{align}

\subsection*{Solutions for $\theta^{(0)(1)}$, $C_i^{(0)(2)}$ and $C_i^{(2)(1)}$ }

The EOM for $\theta^{(0)(1)}$ can be solved analytically. With vanishing boundary conditions imposed at the AdS boundaries, the final solution is given by
\begin{align}
    \theta^{(0)(1)}(r) = & \frac{\partial_v \vartheta_2}{4r_h} \left[ \pi - 2 \arctan\left( \frac{r}{r_h} \right) + \log \frac{r^4 -r_h^4}{r^4} + \log \frac{r+r_h}{r-r_h} \right] \nonumber \\
    & + \frac{\partial_v \vartheta_a}{8\pi r_h} \left[ (2-i) \pi + 2 i \arctan\left( \frac{r}{r_h} \right) - i \log \frac{r+r_h}{r-r_h} \right] \log \frac{r^4 -r_h^4}{r^4}  \label{theta01}
\end{align}
Immediately, we can read off the normalizable modes of $\theta^{(0)(1)}$
\begin{align}
    \tilde O_1^{(0)(1)} = \frac{1}{8} r_h^3 \partial_v \vartheta_a - \frac{1}{4} r_h^3 \partial_v \vartheta_r, \qquad \qquad  \tilde O_2^{(0)(1)} = - \frac{1}{8} r_h^3 \partial_v \vartheta_a - \frac{1}{4} r_h^3 \partial_v \vartheta_r.
\end{align}
From this result, the relevant part in the boundary effective action is given by
\begin{align}
    2m^2 \vartheta \tilde O^{(0)(1)} \bigg|_2^1 = -m^2 r_h^3 \vartheta_a \partial_v \vartheta_r,
\end{align}
which is exactly Eq.~(3.23) in \cite{Glorioso:2018mmw}.

Similarly, the EOM for $C_i^{(0)(2)}$ can be analytically solved. Here, we just record the result
\begin{align}
    C_i^{(0)(2)}(r) = & \frac{\partial_v B_{2i}}{4r_h} \left[ \pi - 2 \arctan\left( \frac{r}{r_h} \right) + 2\log(r+r_h) - \log (r^2 +r_h^2) \right] \nonumber \\
    & - \frac{\partial_v B_{ai}}{8\pi r_h} \left[ -(2-i) \pi - 2 i \arctan\left( \frac{r}{r_h} \right) - i \log \frac{r-r_h}{r+r_h}  \right] \log \frac{r^2 -r_h^2}{r^2 + r_h^2}. \label{Ci02}
\end{align}
So, the normalizable modes of $C_i^{(0)(2)}$ are
\begin{align}
    \tilde J_{1i}^{(0)(2)} = \frac{r_h}{4} \partial_v B_{ai} - \frac{r_h}{2} \partial_v B_{ri}, \qquad \qquad \tilde J_{2i}^{(0)(2)} = - \frac{r_h}{4} \partial_v B_{ai} - \frac{r_h}{2} \partial_v B_{ri}.
\end{align}
Then, we obtain the following pieces in the boundary effective action
\begin{align}
    & B_k \tilde J^{k(0)(2)}\bigg |_2^1 = -r_h B_{ak} \partial_v B_{rk}, \qquad  \frac{1}{2}m^2 B_k \tilde J^{k(0)(2)} \bigg |_2^1 = - \frac{1}{2} m^2 r_h B_{ak} \partial_v B_{rk},
\end{align}
where the first equality exactly reproduces the $v_{00}$-term in \cite{Glorioso:2018mmw}, see equations (4.6) and (4.69) therein.

We turn to $C_i^{(2)(1)}$, which is more involved due to presence of $\theta^{(0)(0)}$ in the source $j_i^{(2)(1)}$, \textit{cfr.} \eqref{source_terms}. The final result is
\begin{align}
    C_i^{(2)(1)}(r) =& - \frac{i B_{ai}}{24\pi} \left[ \pi^2 + 3 \log^2(2r_h^2) + 3 \log\frac{r^2- r_h^2}{4r_h^4} \log(r^2+ r_h^2) + 6 {\rm Li}_2\left( \frac{1}{2} - \frac{r^2}{2r_h^2} \right) \right] \nonumber \\
    & + \frac{1}{4} B_{2i} \log(r^2+r_h^2)  - \frac{1}{4} \partial_i \vartheta_2 \log(r^2+ r_h^2) + \frac{i }{48 \pi} \partial_i \vartheta_a \left[ \pi^2 -12 i\pi \log\frac{r}{r_h} \right. \nonumber \\
    & \left. -6 \log\frac{r^2}{r_h^2} \log\left(\frac{r^2}{r_h^2} -1 \right) - 12 \log r \log\left( \frac{r^2}{r_h^2} +1 \right) + 6 \log \left( r^2+r_h^2\right) \log \left( \frac{r^2}{r_h^2} -1\right) \right. \nonumber \\
    & \left. -3 {\rm Li}_2 \left( \frac{r^4}{r_h^4} \right) \right], \label{Ci21}
\end{align}
where ${\rm Li}_2(x)$ is a polylogarithm function, which has a branch cut point at $x=1$. The normalizable modes of $C_i^{(2)(1)}(r)$ are
\begin{align}
    &\tilde J_{1i}^{(2)(1)} = \left[ \frac{r_h^2}{8} - \frac{ir_h^2}{2\pi} + \frac{i r_h^2}{2\pi} \log(2r_h^2) \right] B_{ai} + \frac{r_h^2}{4} B_{ri} - \left(\frac{r_h^2}{8} + \frac{i r_h^2 \log r_h}{2\pi} \right) \partial_i \vartheta_a - \frac{r_h^2}{4} \partial_i \vartheta_r, \nonumber \\
    &\tilde J_{2i}^{(2)(1)} = \left[ - \frac{r_h^2}{8} - \frac{ir_h^2}{2\pi} + \frac{i r_h^2}{2\pi} \log(2r_h^2) \right] B_{ai} + \frac{r_h^2}{4} B_{ri} + \left(\frac{r_h^2}{8} - \frac{i r_h^2 \log r_h}{2\pi} \right) \partial_i \vartheta_a - \frac{r_h^2}{4} \partial_i \vartheta_r,
\end{align}
from which we obtain the following part in the boundary effective action
\begin{align}
     m^2 B_k \tilde J_k^{(2)(1)} \bigg|_2^1 =& m^2 \frac{r_h^2}{4} B_{rk} (B_{ak} - \partial_k \vartheta_a) + m^2 \frac{r_h^2}{4} B_{ak} (B_{rk} - \partial_k \vartheta_r) \nonumber \\
     &- m^2 \frac{i r_h^2}{2\pi} \left[ 1- \log(2r_h^2) \right] B_{ak} B_{ak}   - m^2 \frac{ir_h^2 \log r_h}{2\pi} B_{ak} \partial_k \vartheta_a.
  \end{align}

\subsection*{The computation of $C_v^{(2)(2)}$ and $\tilde J_v^{(2)(2)}$}

The knowledge of $C_v^{(2)(2)}$ will yield $B_v \partial_v B_v$ and $B_v \partial_v^2 \vartheta$. Here, we are more interested in the first structure $B_v \partial_v B_v$, so that it is sufficient to set $\vartheta_1 = \vartheta_2 =0$ when solving $C_v^{(2)(2)}$. Then, we are able to directly solve the EOM for $C_v^{(2)(2)}$
\begin{align}
    \partial_r \left[ r^3 \partial_r C_v^{(2)(2)}(r) \right] = j_v^{(2)(2)}(r).
\end{align}
We skip the expression for $C_v^{(2)(2)}$ but just record the final result for the normalizable modes
\begin{align}
    \tilde J_{sv}^{(2)(2)} = 0 + \mathcal{O}(\partial_v^2 \vartheta),
\end{align}
which means at the order $\mathcal{O}(m^2)$, the term $B_{av} \partial_v B_{rv}$ again has a zero coefficient\footnote{But see the discussion in section \ref{sec:hol_results}.}
\begin{align}
    m^2 B_v \tilde J^{v(2)(2)} \bigg|_2^1 =0.
\end{align}

\subsection*{Higher order corrections to $\theta$ and $C_i$: Green's function approach}

For the higher order corrections $\theta^{(0)(2)}$ and $C_i^{(2)(2)}$, the source terms in their EOMs turn out to be very lengthy. Therefore, we will solve them via the Green's function approach.

We turn to the construction of the Green's functions based on the linearly independent solutions for homogeneous EOMs. The Green's functions $G_X(r,r^\prime)$ and $G_Y(r,r^\prime)$ obey
\begin{align}
    \partial_r\left[r^3f(r) G_X(r,r^\prime) \right] = \delta(r-r^\prime), \qquad \qquad  \partial_r\left[r^5f(r) G_Y(r,r^\prime) \right] = \delta(r-r^\prime).
\end{align}
Since higher order corrections $C_i^{(2)(2)}$, $\theta^{(0)(2)}$ satisfy vanishing-type Dirichlet boundary conditions,\footnote{We have subtracted the leading terms near the boundary, as read off from the general analysis \eqref{Cmu_theta_bdy}.} we require the Green's functions to satisfy similar conditions
\begin{align}
    &G_X(r \to \infty_2, r^\prime) = 0 + \frac{\#_X^2}{r^2}, \qquad G_X(r \to \infty_1, r^\prime) = 0 + \frac{\#_X^1}{r^2}, \nonumber \\
    & G_Y(r \to \infty_2, r^\prime) = 0 + \frac{\#_Y^2}{r^4}, \qquad G_Y(r \to \infty_1, r^\prime) = 0 + \frac{\#_Y^1}{r^4}.
\end{align}
Accordingly, we make linear combinations over the linearly independent solutions in Eq.~\eqref{linear_indep_solutions} to get
\begin{align}
   & X_2(r) = -\frac{1}{2} \log \frac{r^2-r_h^2}{r^2 +r_h^2},  \qquad \qquad X_1(r) = -\frac{1}{2} \log \frac{r^2-r_h^2}{r^2 +r_h^2} + i \pi, \nonumber \\
   & Y_2(r) = -\frac{1}{2} \log \frac{r^4-r_h^4}{r^4},  \qquad \qquad \, \, Y_1(r) = -\frac{1}{2} \log \frac{r^4-r_h^4}{r^4} + i \pi,
\end{align}
which satisfy ``good'' boundary conditions near the AdS boundaries:
\begin{align}
    &X_2(r \to \infty_2) = 0 + \frac{r_h^2}{r^2} + \cdots, \qquad \qquad \,\, X_2(r \to \infty_1) = -i\pi + \frac{r_h^2}{r^2} + \cdots, \nonumber \\
    &X_1(r \to \infty_2) = i\pi + \frac{r_h^2}{r^2} + \cdots, \qquad \qquad X_1(r \to \infty_1) = 0 + \frac{r_h^2}{r^2} + \cdots,
\end{align}
and
\begin{align}
    &Y_2(r \to \infty_2) = 0 + \frac{r_h^4}{2r^4} + \cdots, \qquad \qquad \, \, Y_2(r \to \infty_1) = - i\pi + \frac{r_h^4}{2r^4} + \cdots, \nonumber \\
    &Y_1(r \to \infty_2) = i\pi + \frac{r_h^4} {2r^4} + \cdots, \qquad \qquad Y_1(r \to \infty_1) = 0 + \frac{r_h^4}{2r^4} + \cdots,
\end{align}
Now, we are ready to build the Green's functions based on the new set of linearly independent solutions above
\begin{align}
    &G_X(r,r^\prime) = \frac{1}{r^{\prime3}f(r^\prime) W_X(r^\prime)} \left[ \Theta(r-r^\prime) X_1(r) X_2(r^\prime) + \Theta(r^\prime - r) X_2(r) X_1(r^\prime) \right], \nonumber \\
    & G_Y(r,r^\prime) = \frac{1}{r^{\prime5}f(r^\prime) W_Y(r^\prime)} \left[ \Theta(r-r^\prime) Y_1(r) Y_2(r^\prime) + \Theta(r^\prime - r) Y_2(r) Y_1(r^\prime) \right],
\end{align}
where $W_X$ and $W_Y$ are the Wronskian determinants of solutions $\{ X_1, X_2\}$ and $\{Y_1, Y_2\}$, respectively
\begin{align}
   & W_X(r) \equiv X_2(r) \partial_r X_1(r) - X_1(r) \partial_r X_2(r) = \frac{2i\pi r_h^2}{r^3f(r)}, \nonumber \\
    &W_Y(r) \equiv Y_2(r) \partial_r Y_1(r) - Y_1(r) \partial_r Y_2(r) = \frac{2i\pi r_h^4}{r^5f(r)}.
\end{align}
The function $\Theta(r-r^\prime)$ is a step function compatible with the radial contour. Apparently, near the AdS boundary, thus-constructed Green's functions behave as
\begin{align}
    &G_X(r\to \infty_2, r^\prime) = 0 +  \frac{X_1(r^\prime)}{2i\pi} \frac{1}{r^2} + \cdots, \qquad \qquad  G_X(r\to \infty_1, r^\prime) = 0 +  \frac{X_2(r^\prime)}{2i\pi} \frac{1}{r^2} + \cdots, \nonumber \\
    &G_Y(r\to \infty_2, r^\prime) = 0 +  \frac{Y_1(r^\prime)}{2i\pi} \frac{1}{2r^4} + \cdots, \qquad \qquad  G_Y(r\to \infty_1, r^\prime) = 0 +  \frac{Y_2(r^\prime)}{2i\pi} \frac{1}{2r^4} + \cdots\,.
\end{align}

Then, the solutions for $C_i^{(2)(2)}$ and $\theta^{(0)(2)}$ are
\begin{align}
   C_i^{(2)(2)}(r) & = \int_{\infty_2}^{\infty_1} dr^\prime G_X(r,r^\prime) j_i^{(2)(2)} (r^\prime) \nonumber \\
   & = \frac{X_1(r)}{2i\pi r_h^2} \int_{\infty_2}^r dr^\prime X_2(r^\prime) j_i^{(2)(2)} (r^\prime) + \frac{X_2(r)}{2i\pi r_h^2} \int_r^{\infty_1} dr^\prime X_1(r^\prime) j_i^{(2)(2)}(r^\prime), \nonumber \\
   \theta^{(0)(2)}(r) & = \int_{\infty_2}^{\infty_1} dr^\prime G_Y(r,r^\prime)  j_\theta^{(0)(2)}(r^\prime) \nonumber \\
   & = \frac{Y_1(r)}{2i\pi r_h^4} \int_{\infty_2}^r dr^\prime Y_2(r^\prime) j_\theta^{(0)(2)} (r^\prime) + \frac{Y_2(r)}{2i\pi r_h^4} \int_r^{\infty_1} dr^\prime Y_1(r^\prime) j_\theta^{(0)(2)}(r^\prime). \label{Ci22_theta02_GreenFunction}
\end{align}
In order to extract the normalizable modes in $C_i^{(2)(2)}$ and $\theta^{(0)(2)}$, we take the near-boundary limits $r \to \infty_{1,2}$ of \eqref{Ci22_theta02_GreenFunction}
\begin{align}
    C_i^{(2)(2)}(r\to \infty_1) = & \frac{X_1(r)}{2i\pi r_h^2} \int_{\infty_1}^r dr^\prime X_2(r^\prime) j_i^{(2)(2)} (r^\prime) - \frac{X_2(r)}{2i\pi r_h^2} \int_{\infty_1}^r dr^\prime X_1(r^\prime) j_i^{(2)(2)}(r^\prime) \nonumber \\
    & + \frac{X_1(r)}{2i\pi r_h^2} \int_{\infty_2}^{\infty_1} dr^\prime X_2(r^\prime) j_i^{(2)(2)}(r^\prime), \nonumber \\
    C_i^{(2)(2)}(r\to \infty_2) = & \frac{X_1(r)}{2i\pi r_h^2} \int_{\infty_2}^r dr^\prime X_2(r^\prime) j_i^{(2)(2)} (r^\prime) - \frac{X_2(r)}{2i\pi r_h^2} \int_{\infty_2}^r dr^\prime X_1(r^\prime) j_i^{(2)(2)}(r^\prime) \nonumber \\
    & + \frac{X_2(r)}{2i\pi r_h^2} \int_{\infty_2}^{\infty_1} dr^\prime X_1(r^\prime) j_i^{(2)(2)}(r^\prime), \nonumber \\
    \theta^{(0)(2)}(r\to \infty_1) = & \frac{Y_1(r)}{2i\pi r_h^4} \int_{\infty_1}^r dr^\prime Y_2(r^\prime) j_\theta^{(0)(2)} (r^\prime) - \frac{Y_2(r)}{2i\pi r_h^4} \int_{\infty_1}^r dr^\prime Y_1(r^\prime) j_\theta^{(0)(2)}(r^\prime) \nonumber \\
    & + \frac{Y_1(r)}{2i\pi r_h^4} \int_{\infty_2}^{\infty_1} dr^\prime Y_2(r^\prime) j_\theta^{(0)(2)}(r^\prime), \nonumber \\
    \theta^{(0)(2)}(r\to \infty_2) = & \frac{Y_1(r)}{2i\pi r_h^4} \int_{\infty_2}^r dr^\prime Y_2(r^\prime) j_\theta^{(0)(2)} (r^\prime) - \frac{Y_2(r)}{2i\pi r_h^4} \int_{\infty_2}^r dr^\prime Y_1(r^\prime) j_\theta^{(0)(2)}(r^\prime) \nonumber \\
    & + \frac{Y_2(r)}{2i\pi r_h^4} \int_{\infty_2}^{\infty_1} dr^\prime Y_1(r^\prime) j_\theta^{(0)(2)}(r^\prime), \label{Ci22_theta02_AdS}
\end{align}
Indeed, in each equality above, near the AdS boundary $r^\prime = \infty_1$ or $r^\prime = \infty_2$, the integrals diverge, but the final sum is free of such divergences. For safety, we introduce a regulator, $r^\prime = \Lambda_s$, near the two AdS boundaries.

We would like to stress that we are interested in extracting the large $r$ behavior for $C_i^{(2)(2)}$ and $\theta^{(0)(2)}$. The contour integrals in \eqref{Ci22_theta02_AdS} will be computed by splitting the radial contour as
\begin{align}
    \int_{\infty_2}^{\infty_1} dr = \int_{\infty_2}^{r_h + \epsilon} dr + \int_{\mathcal C} i \epsilon e^{i\varphi} d\varphi + \int_{r_h +\epsilon}^{\infty_1} dr,
\end{align}
where $\mathcal C$ denotes the infinitesimal circle, and the angle $\varphi$ runs from $0$ to $2\pi$, as one goes from upper branch to lower branch. Here, it is important to note that as $\epsilon \to 0$ is taken, the contribution from the circle does not vanish, and moreover brings in $\log \epsilon$-divergence. The contributions from upper and lower legs cancel significantly. Below, we present the respective results for the gauge sector and scalar sector.\\

{\bf $\bullet$ The results for the normalizable modes $\tilde J_i^{(2)(2)}$.} The correction at this order yields to the terms $\partial_v B_i$, $\partial_v \partial_i \vartheta$ in the normalizable modes $\tilde J_i^{(2)(2)}$. For simplicity, we will not capture $\partial_v \partial_i \vartheta$-terms so that we can simply drop the $\theta$-term in the source $j_i^{(2)(2)}$ of \eqref{source_terms}. To this end, we have collected the compact expressions for $C_i^{(0)(2)}$ and $C_i^{(2)(1)}$ in \eqref{Ci02} and \eqref{Ci21}. Skipping the details, we present the final results
\begin{align}
    &\tilde J_{1i}^{(2)(2)} = \left[ -\frac{r_h}{8} + \frac{1}{8} r_h \log(2r_h^2) \right] \partial_v B_{ai}
    - \frac{1}{4}r_h \left[ -1 + \log(2r_h^2) \right] \partial_v B_{ri}, \nonumber \\
    &\tilde J_{2i}^{(2)(2)} = \left[ \frac{r_h}{8} - \frac{1}{8} r_h \log(2r_h^2) \right] \partial_v B_{ai}
    - \frac{1}{4}r_h \left[ -1 + \log(2r_h^2) \right] \partial_v B_{ri}.
\end{align}
The relevant terms in the effective action are
\begin{align}
    m^2 B_k \tilde J^{k(2)(2)} \bigg |_2^1 = m^2 \frac{r_h}{2} \left[ 1- \log(2r_h^2) \right] B_{ak} \partial_v B_{rk}.
\end{align}

{\bf $\bullet$ The results for the normalizable modes $\tilde O^{(0)(2)}$.} The result at this order yields terms $\vartheta \partial_v^2 \vartheta$, $\vartheta \partial_k^2 \vartheta$, $\vartheta \partial_v B_v$, $\vartheta \partial_k B_k$ in the boundary effective action.
The source $j_\theta^{(0)(2)}$ of \eqref{source_terms} involves $\theta^{(0)(0)}$, $\theta^{(0)(1)}$, $C_\mu^{(0)(1)}$, which have been presented in \eqref{theta00}, \eqref{theta01}, \eqref{Cv01} and \eqref{Ci01}.

The calculations are in parallel with those of $C_i^{(2)(2)}$, although more tedious. Here, we just present the final results. For $\tilde O_1^{(0)(2)}$, we have a linear combination of the following terms with coefficients given by
\begin{align}
    &\partial_v B_{rv}: \qquad \frac{r_h^2}{16} \left( -3 + 2 \log 2 + 4 \log r_h \right), \nonumber \\
    &\partial_v B_{av}: \qquad \frac{r_h^2}{32} \left( -3 + 2\log 2 + 4 \log r_h \right) + \frac{i r_h^2}{16\pi} \left( 2\log2 - \log^2 2\right), \nonumber \\
    &\partial_k B_{rk}: \qquad \frac{1}{8}r_h^2, \nonumber \\
    &\partial_k B_{ak}: \qquad \frac{r_h^2}{16} - \frac{i r_h^2}{16\pi} (3- 4 \log r_h), \nonumber \\
    &\partial_v^2 \vartheta_r: \qquad \frac{r_h^2}{8}(1-\log 2), \nonumber \\
    &\partial_v^2 \vartheta_a: \qquad \frac{r_h^2}{16}(1-\log 2) - \frac{ir_h^2}{96\pi} \left( 5\pi^2 +24 \log2 -12 \log^22 \right), \nonumber \\
    &\partial_k^2 \vartheta_r: \qquad - \frac{r_h^2}{8}, \nonumber \\
    &\partial_k^2 \vartheta_a: \qquad - \frac{r_h^2}{16} + \frac{i r_h^2}{4\pi} \log2.
\end{align}
The results for $\tilde O_2^{(0)(2)}$ are quite similar
\begin{align}
    &\partial_v B_{rv}: \qquad \frac{r_h^2}{16} \left( -3 + 2 \log 2 + 4 \log r_h \right), \nonumber \\
    &\partial_v B_{av}: \qquad \frac{r_h^2}{32} \left( 3 - 2 \log 2 - 4 \log r_h \right) + \frac{i r_h^2}{16\pi} \left(2\log2 - \log^2 2 \right), \nonumber \\
    &\partial_k B_{rk}: \qquad \frac{1}{8}r_h^2, \nonumber \\
    &\partial_k B_{ak}: \qquad -\frac{r_h^2}{16} - \frac{i r_h^2}{16\pi} (3- 4 \log r_h), \nonumber \\
    &\partial_v^2 \vartheta_r: \qquad \frac{r_h^2}{8}(1-\log 2), \nonumber \\
    &\partial_v^2 \vartheta_a: \qquad -\frac{r_h^2}{16}(1-\log 2) - \frac{ir_h^2}{96\pi} \left( 5\pi^2 +24 \log2 -12 \log^22 \right), \nonumber \\
    &\partial_k^2 \vartheta_r: \qquad - \frac{r_h^2}{8}, \nonumber \\
    &\partial_k^2 \vartheta_a: \qquad  \frac{r_h^2}{16} + \frac{i r_h^2}{4\pi} \log2.
\end{align}
Thus, the relevant terms in the effective action are
\begin{align}
    & 2m^2 \vartheta \tilde O^{(0)(2)} \bigg |_2^1 \nonumber \\
    = & 2m^2 \frac{r_h^2}{16} \left( 3 - 2\log 2 -4 \log r_h \right) (B_{av} \partial_v \vartheta_r - \vartheta_a \partial_v B_{rv}) \nonumber \\
    & + 2m^2 \frac{-r_h^2}{8} (B_{ak} \partial_k \vartheta_r - \vartheta_a \partial_k B_{rk}) \nonumber \\
    & + 2m^2 \frac{r_h^2}{4} (1-\log 2) \vartheta_a \partial_v^2 \vartheta_r \nonumber \\
    & - 2m^2 \frac{r_h^2}{4} \vartheta_a \partial_k^2 \vartheta_r \nonumber \\
    & + 2m^2 \frac{i r_h^2}{16\pi} (2\log2 - \log^2 2) \vartheta_a \partial_v B_{av} \nonumber \\
    & + 2m^2 \frac{-i r_h^2}{16\pi} (3-4\log r_h) \vartheta_a \partial_k B_{ak} \nonumber \\
    & + 2m^2 \frac{-i r_h^2}{96\pi} (5\pi^2 + 24 \log 2 -12 \log^22) \vartheta_a \partial_v^2 \vartheta_a \nonumber \\
    & + 2m^2 \frac{ir_h^2 \log 2}{4\pi} \vartheta_a \partial_k^2 \vartheta_a, \label{vartheta_O02}
\end{align}
where the fourth line and fifth line are in perfect agreement with Eq.(3.29) of \cite{Glorioso:2018mmw}, and all the rest terms are new. The expressions (in terms of contour integrals) for the last two lines have been presented in \cite{Glorioso:2018mmw}, without explicit values.

\bibliographystyle{JHEP}
\bibliography{biblio.bib}
\end{document}